\documentclass[10pt]{iopart}
\usepackage{epsfig}
\usepackage{supertabular}
\usepackage{setspace}
\usepackage[authoryear]{natbib}
\begin{document}
\title{Is high-frequency trading inducing changes in market microstructure and dynamics?}
\author{Reginald Smith}
\address{PO Box 10051, Rochester, NY 14610}
\ead{rsmith@bouchet-franklin.org}

\begin{abstract}
Using high-frequency time series of stock prices and share volumes sizes from January 2002-May 2009, this paper investigates whether the effects of the onset of high-frequency trading, most prominent since 2005, are apparent in the dynamics of the dollar traded volume. Indeed it is found in almost all of 14 heavily traded stocks, that there has been an increase in the Hurst exponent of dollar traded volume from Gaussian noise in the earlier years to more self-similar dynamics in later years. This shift is linked both temporally to the Reg NMS reforms allowing high-frequency trading to flourish as well as to the declining average size of trades with smaller trades showing markedly higher degrees of self-similarity.
\end{abstract}
\maketitle
\section{Introduction}
Over the last couple of decades, cheap computing power and improved telecommunications have all but obsoleted to old style open-cry auctions which drove equity markets for centuries. This change has occurred so rapidly, that it has disrupted the traditional dominance of the exchanges, brought into finance new strategies such as statistical arbitrage, and fuelled intense public and regulatory debate over the positive benefits and costs of the new world of electronic trading.

One of the largest effects of this movement has been the disruption of and controversy about old paradigms such as assumed normal distributions of asset returns and the absence of long-term persistence in price fluctuations that random walk models dictate. However, recent advances in data analysis techniques and recent financial market turbulence may provide a fertile proving ground for research that improves both a theoretical and empirical understanding of financial market dynamics.

The field of research which has likely attracted some of the largest attention and literature in quantitative finance and econophysics \cite{econo1, econo2} is likely the research of the movements by asset prices and volumes in trading on financial markets. This is for several reasons. First, the data, which are fundamentally time series amongst correlated systems, are amenable to the well-developed techniques from econometrics, physical research, and applied mathematics and statistics. Second, large datasets are easily and cheaply available, sometimes for free, as in the case of daily closing prices. Combined with cheap computing power, there is a low barrier to entry for participants eager to apply many well-known techniques to price dynamics of stocks, currencies, or other widely traded assets. 

A common theme in this research has often been the investigation of self-similarity and long-range dependence in the time series. The former is often measured by the Hurst exponent, $H$, which measures the relative degree of self-similarity from pure Markovian Browninan motion at $H=0.5$ to complete self-similarity at $H=1$. A value of $0.5 < H < 1$ is typically described by fractional Brownian motion which demonstrates self-similarity at multiple time scales of fluctuations in contrast to the independent fluctuations of Brownian motion. Self-similarity and long-range dependence in financial time series has a long and sometimes contentious history. There are many papers disputing whether the most common measure, log asset returns, display any sort of self-similarity which would bring the efficient market hypothesis into questions \cite{price1,price2,price3} Studies on trading volume \cite{volume1, volume2} or traded value seem tend to demonstrate $H>0.5$ over long time scales. One key exception, however, tends to be short term intraday equity trading data. In particular, in \cite{intraday1,intraday2,intraday3} intraday trading dollar value, on the order of times less than 60 minutes, universally demonstrate $H \approx 0.5$ indicating there exists different scaling behavior at different time scales for equity trading volumes. For long periods of intraday trading or trades across multiple days, there is almost always a significant deviation from Gaussian noise and $H > 0.5$. Therefore, traditionally the correlation structure of trading time series is generated at time scales of the orders of many minutes or hours and is not an inherent feature of the dynamics at shorter timescales.

This paper is organized as follows. First, a brief introduction to the history of financial markets in the in the US will be given which culminates in the reforms of the late 20th and early 21st century which enabled high frequency trading. Next, the data sources and mathematical methods used to analyze the time series self-similarity will be introduced and explained. Finally the results of the analysis, possible causes, and a brief conclusion will close out the paper.

\section{A brief history of the events leading up to high frequency trading}
\begin{table*}[!tc] \vspace{1.5ex}
\centering
\begin{tabular}{|c|c|}
\hline
Acronym & Meaning\\
\hline
ATS&Alternative trading system\\
ECN&Electronic communication network\\
HFT&High-frequency trading\\
NASD&National Association of Securities Dealers\\
NASDAQ&National Association of Securities Dealers Automated Quotations\\
NBBO&National Best Bid and Offer\\
NYSE&New York Stock Exchange\\
OTC&Over-the-counter\\
Reg ATS&Regulation Alternative Trading System\\
Reg NMS&Regulation National Market System\\
SEC&Securities \& Exchange Commission\\
TAQ&NYSE Trades and Quotes Database\\
\hline
\end{tabular}
\caption{Acronym Guide}
\end{table*}
In 1792, as a new nation worked out its new constitution and laws, another less heralded revolution began when several men met under a Buttonwood tree, and later coffee houses, on Wall St. in New York City to buy and sell equity positions in fledgling companies. An exclusive members club from the beginning, the New York Stock Exchange (NYSE) rapidly grew to become one of the most powerful exchanges in the world. Ironically, even the non-member curbside traders outside the coffee houses gradually evolved into over-the-counter (OTC) traders and later, the NASDAQ. A very detailed and colorful history of this evolution is given in \cite{markethistory1,markethistory2}. 

Broadly, the role of the exchange is to act as a market maker for company stocks where buyers with excess capital would like to purchase shares and sellers with excess stock would like to liquidate their position. Several roles developed in the NYSE to encourage smooth operation and liquidity. There came to be several types of market makers for buyers and sellers known as brokers, dealers, and specialists. The usual transaction involves the execution of a limit order versus other offers. A limit order, as contrasted to a market order which buys or sells a stock at the prevailing market rate, instructs the purchase of a stock up to a limit ceiling or the sale of a stock down to a limit floor. 
Brokers act on behalf of a third-party, typically an institutional investor, to buy or sell stock according to the pricing of the limit order. Dealers, also known as market-makers, buy and sell stock using their own capital, purchasing at the bid price and selling at the ask price, pocketing the bid-ask spread as profit. This helps ensure market liquidity. A specialist acts as either a broker or dealer but only for a specific list of stocks that he or she is responsible for. As a broker, the specialist executes trades from a display book of outstanding orders and as a dealer a specialist can trade on his or her own account to stabilize stock prices.

The great rupture in the business-as-usual came with the Great Depression and the unfolding revelations of corrupt stock dealings, fraud, and other such malfeasance. The Securities and Exchange Commission (SEC) was created by Congress in 1934 by the Securities Exchange Act. Since then, it has acted as the regulator of the stock exchanges and the companies that list on them. Over time, the SEC and Wall Street have evolved together, influencing each other in the process.

By the 1960s, the volume of traded shares was overwhelming the traditional paper systems that brokers, dealers, and specialists on the floor used. A``paperwork crisis'' developed that seriously hampered operations of the NYSE and led to the first electronic order routing system, DOT by 1976. In addition, inefficiencies in the handling of OTC orders, also known as ``pink sheets'', led to a 1963 SEC recommendation of changes to the industry which led the National Association of Securities Dealers (NASD) to form the NASDAQ in 1968. Orders were displayed electronically while the deals were made through the telephone through``market makers'' instead of dealers or specialists. In 1975, under the prompting of Congress, the SEC passed the Regulation of the National Market System, more commonly known as Reg NMS, which was used to mandate the interconnectedness of various markets for stocks to prevent a tiered marketplace where small, medium, and large investors would have a specific market and smaller investors would be disadvantaged. One of the outcomes of Reg NMS was the accelerated use of technology to connect markets and display quotes. This would enable stocks to be traded on different, albeit connected, exchanges from the NYSE such as the soon to emerge electronic communication networks (ECNs), known to the SEC as alternative trading systems (ATS).

In the 1980s, the NYSE upgraded their order system again to SuperDot. The increasing speed and availability computers helped enable trading of entire portfolios of stocks simultaneously in what became known as program trading. One of the first instances of algorithmic trading, program trading was not high-frequency per se but used to trade large orders of multiple stocks at once. Program trading was profitable but is now often cited as one of the largest factors behind the 1987 Black Monday crash. Even the human systems broke down, however, as many NASDAQ market makers did not answer the phones during the crash.

The true acceleration of progress and the advent of what is now known as high frequency trading occurred during the 1990s. The telecommunications technology boom as well as the dotcom frenzy led to many extensive changes. A new group of exchanges became prominent. These were the ECN/ATS exchanges. Using new computer technology, they provided an alternate market platform where buyers and sellers could have their orders matched automatically to the best price without middlemen such as dealers or brokers. They also allowed complete anonymity for buyers and sellers. One issue though was even though they were connected to the exchanges via Reg NMS requirements, there was little mandated transparency. In other words, deals settled on the ECN/ATS were not revealed to the exchange. On the flip side, the exchange brokers were not obligated to transact with an order displayed from an ECN, even if it was better for their customer.

This began to change, partially because of revelations of multiple violations of fiduciary duty by specialists in the NYSE. One example, similar to the soon to be invented `flash trading', was where they would ``interposition'' themselves between their clients and the best offer in order to either buy low from the client and sell higher to the NBBO (National Best Bid and Offer; the best price) price or vice versa. In 1997, the SEC passed the Limit Order Display rule to improve transparency that required market makers to include offers at better prices than those the market maker is offering to investors. This allows investors to know the NBBO and circumvent corruption. However, this rule also had the effect of requiring the exchanges to display electronic orders from the ECN/ATS systems that were better priced. The SEC followed up in 1998 with Regulation ATS. Reg ATS allowed ECN/ATS systems to register as either brokers or exchanges. It also protected investors by mandating reporting requirements and transparency of aggregate trading activity for ECN/ATS systems once they reach a certain size.

These changes opened up huge new opportunities for ECN/ATS systems by allowing them to display and execute quotes directly with the big exchanges. Though they were required to report these transactions to the exchange, they gained much more. In particular, with their advanced technology and low-latency communication systems, they became a portal through which next generation algorithmic trading and high frequency trading algorithms could have access to wider markets. Changes still continued to accelerate.

In 2000, were two other groundbreaking changes. First was the decimalization of the price quotes on US stocks. This reduced the bid-ask spreads and made it much easier for computer algorithms to trade stocks and conduct arbitrage. The NYSE also repealed Rule 390 which had prohibited the trading of stocks listed prior to April 26, 1979 outside of the exchange. High frequency trading began to grow rapidly but did not truly take off until 2005.

In June 2005, the SEC revised Reg NMS with several key mandates. Some were relatively minor such as the Sub-Penny rule which prevented stock quotations at a resolution less than one cent. However, the biggest change was Rule 611, also known as the Order Protection Rule. Whereas with the Limit Order Display rule, exchanges were merely required to display better quotes, Reg NMS Rule 611 mandated, with some exceptions, that trades are always automatically executed at the best quote possible. Price is the only issue and not counterparty reliability, transaction speed, etc. The opening for high frequency trading here is clear. The automatic trade execution created the perfect environment for high speed transactions that would be automatically executed and not sit in a queue waiting for approval by a market maker or some vague exchange rule. The limit to trading speed and profit was now mostly the latency on electronic trading systems.

\begin{figure}
\centering
\includegraphics[height=2.75in,width=2.75in]{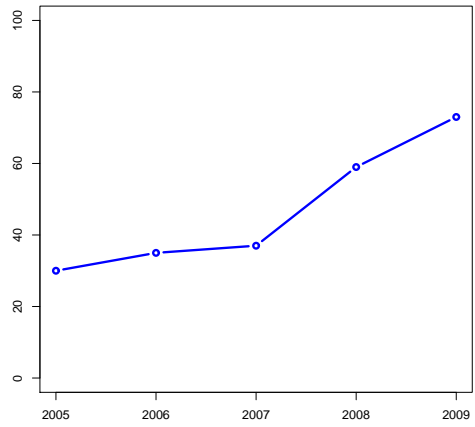}
\caption{The proportion of all US equity trading carried out through high frequency trading. Source: Tabb Group}
\label{tabb}
\end{figure}

The boom in ECN/ATS business created huge competition for exchanges causing traditional exchanges (NYSE \& Euronext) to merge and some exchanges to merge with large ECNs (NYSE \& Archipelago). In addition, the competition created increasingly risky business strategies to lure customers. CBSX and DirectEdge pioneered `flash trading' on the Chicago Board of Exchange and the NYSE/NASDAQ respectively where large limit orders would be flashed for 50 milliseconds on the network to paying customers who could then rapidly trade in order to profit from them before public advertisement. Many of these were discontinued in late 2009 after public outcry but HFT was already the dominant vehicle for US equity trading as shown in figure \ref{tabb}. 

\begin{figure}
\centering
\begin{tabular}{cc}
\includegraphics[height=2.75in,width=2.75in]{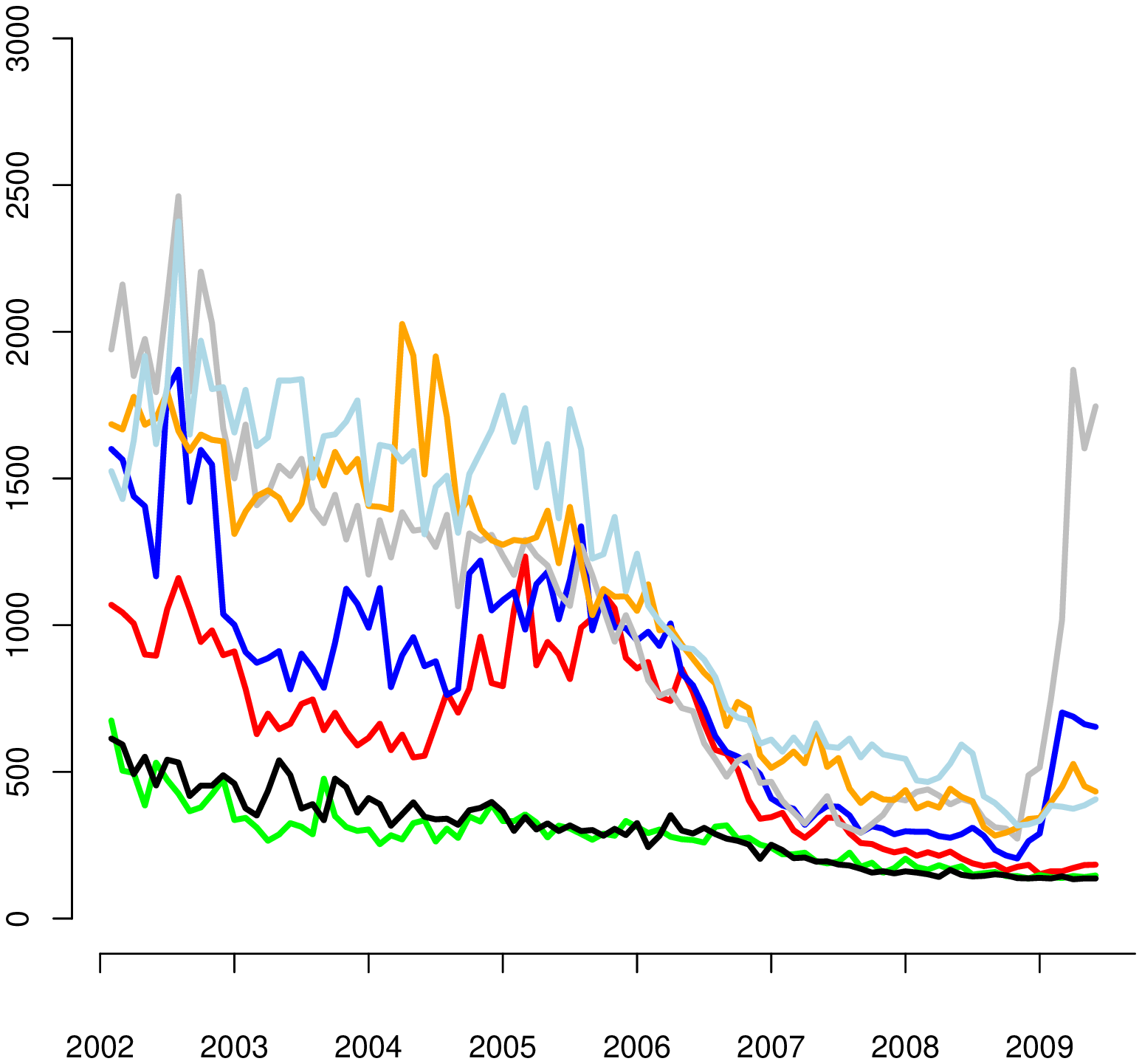} & 
\includegraphics[height=2.75in,width=2.75in]{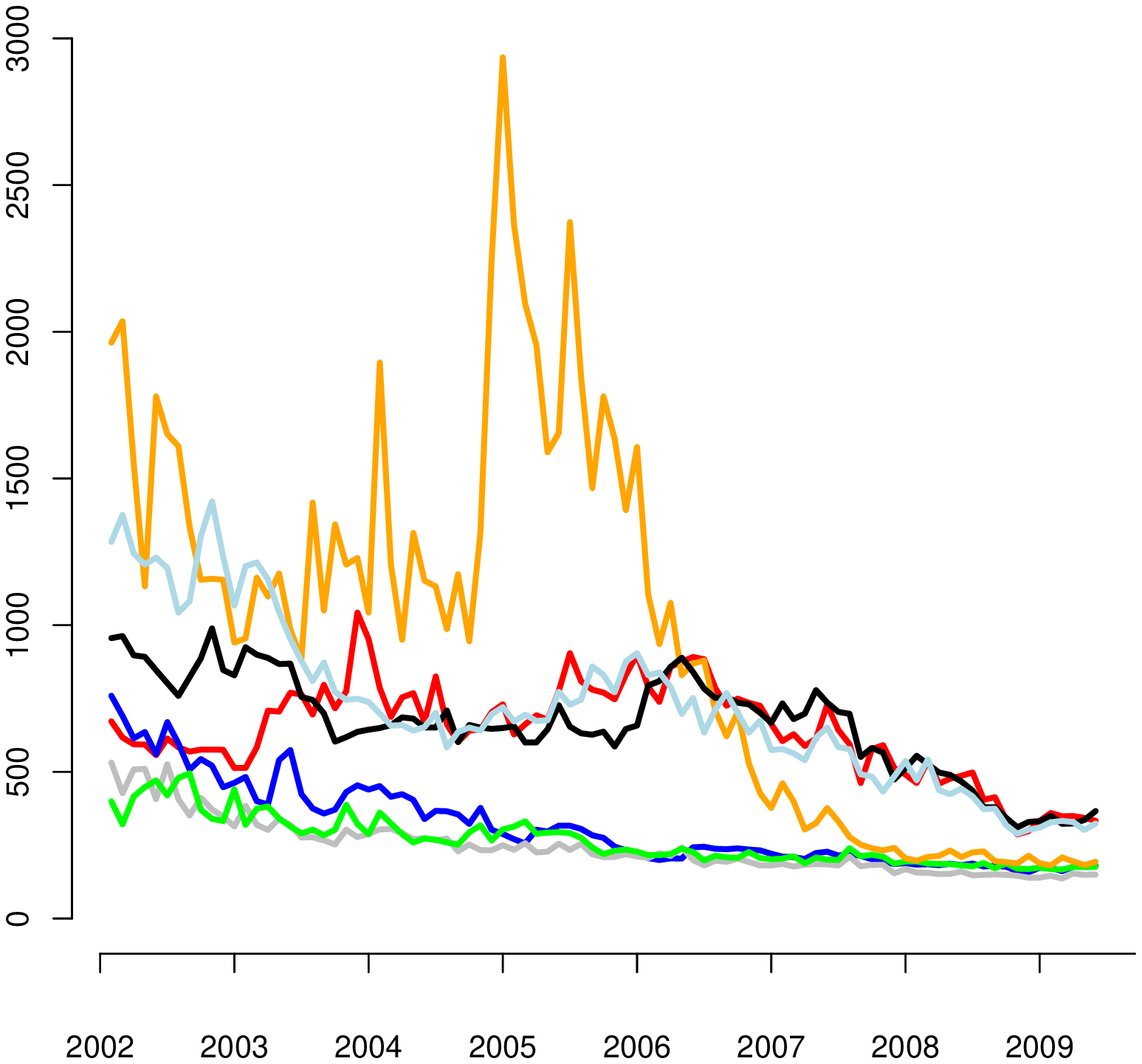} \\
\end{tabular}
\label{sharesize}
\caption{A graph showing the average size of a trade in shares over the period from 2002-May 2009 where data for this paper is used. The left image is for stocks on the NYSE and the right image is for stocks on the NASDAQ. The corresponding stocks for the NYSE/NASDAQ for each color are: red, PG/MSFT, blue, BAC/AAPL, grey, C/GENZ, green, CHD/GILD, orange, GE/NWS, black, ITT/INTC, light blue, PFE/CSCO. The spikes in shares/trade for both Citigroup and Bank of America in early 2009 are due to emergency injections of capital by the US government. In Citigroup's case, this was done by converting bonds held by the US Treasury into stock.}
\end{figure}

\begin{figure}
\centering
\includegraphics[height=5in,width=7in]{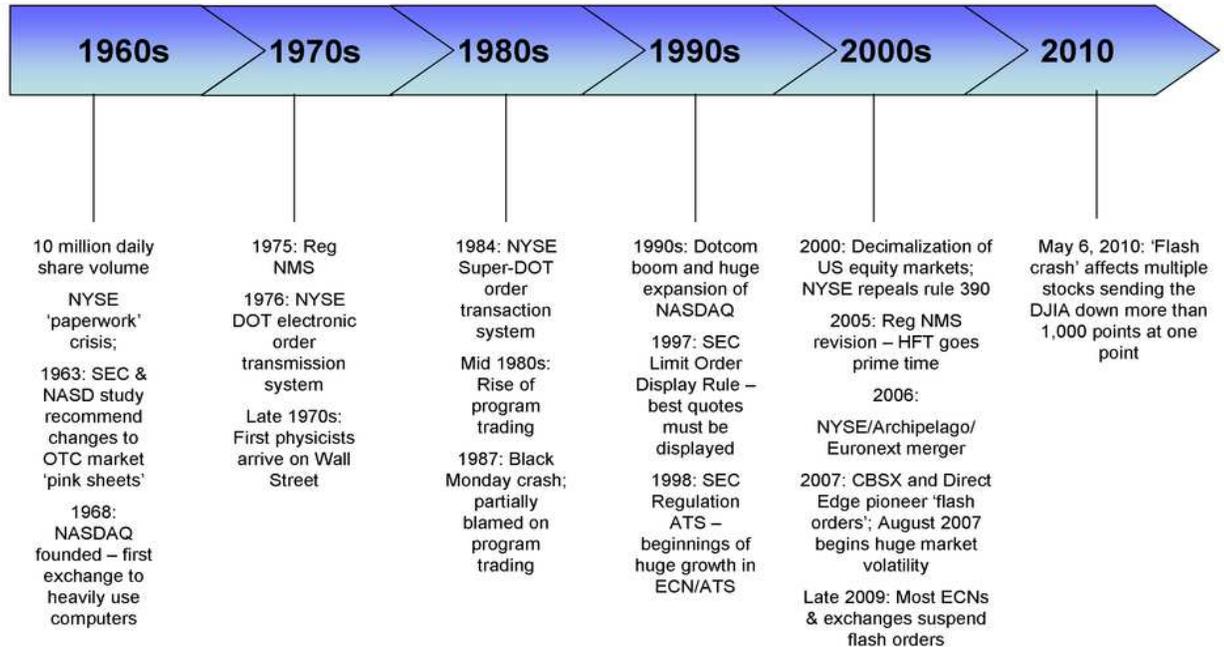}
\caption{The rough timeline of the events spanning decades that led to the current market penetration of high-frequency trading.}
\label{timeline}

\end{figure}

\begin{figure}
\centering
\includegraphics[height=5in,width=7in]{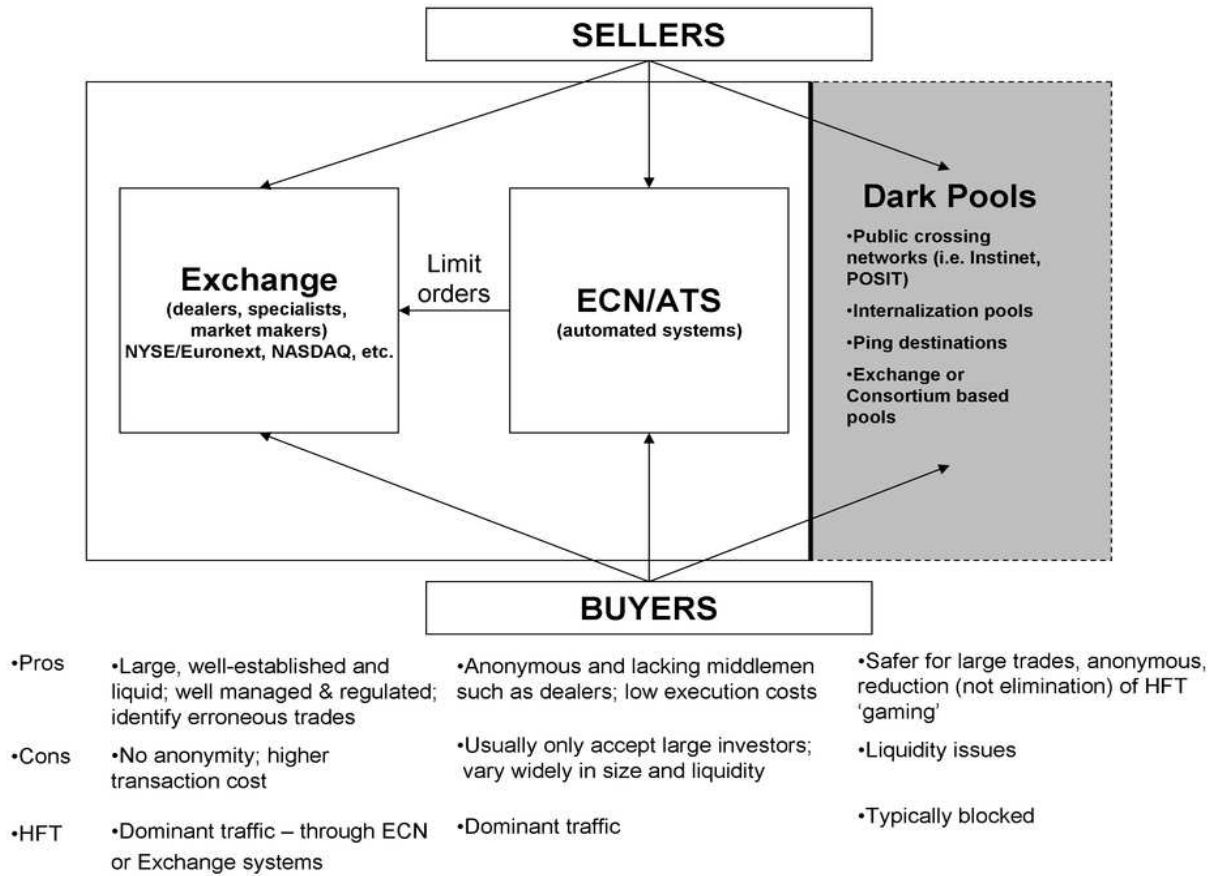}
\caption{A basic overview of the current major venues for the trading of US equities.}
\label{market}

\end{figure}

HFT thrives on rapid fire trading of small sized orders and the overall shares/trade has dropped rapidly over the last few years is shown in figure \ref{sharesize}. In addition, the HFT strategy of taking advantage of pricing signals from large orders has forced many orders off exchanges into proprietary trading networks called `dark pools'  which get their name from the fact they are private networks which only report the prices of transactions after the transaction has occurred and typically anonymously match large orders without price advertisements. These dark pools allow a safer environment for large trades which (usually) keep out opportunistic high frequency traders. The basic structure of today's market and a timeline of developments are given in figure \ref{timeline} and figure \ref{market}. For more detailed information, see \cite{mktinfo1,mktinfo2,mktinfo3,mktinfo4,mktinfo5,mktinfo6}

\section{Data Sources and Analysis Methods}

In this paper, we will follow up on the research previously done in \cite{intraday1,intraday2,intraday3} and describe changes in the high-frequency (short period) structure of the equity trading markets that likely have been induced by the spread of HFT firms and strategies. In particular, this paper will demonstrate, that since 2005 there has been a marked changed in the correlation structure of stock trading dynamics where amongst many stocks, there has been a measurable departure from the typical $H=0.5$ regime and that stronger self-similarity has been steadily increasing over the same period in the time that HFT has become the largest source of market volume.

The data in this paper uses the daily intraday trading records of 14 US stocks, 7 from the NYSE and 7 from the NASDAQ, from the NYSE Trades and Quotes (TAQ) database accessed through Wharton Research Data Services (WRDS). The stocks used are listed in table \ref{stocklist}. These stocks were chosen for being both relatively liquid (high daily trading volumes) and also representing a variety of industries. Time series of all intraday trades occurring from 0930 to 1600 EST were gathered over the period January 1, 2002 to May 31, 2009. The TAQ database is comprehensive in coverage of trades settled through the NYSE and NASDAQ which includes trades transacted by exchanges with ECNs. However, the TAQ data does not cover private ECN transactions not settled with the exchanges nor the information from trades in dark pools. Thus it provides a good, but not complete, view of the market and HFT. In addition, sale condition codes and trade correction indicators corresponding to invalid trades were removed. Trades with sale condition codes of (B, D, G, J, K, L, M, N, O, P, Q, R, T, U, W, Z, 4 (NASDAQ), 6(NASDAQ)), trade correction indicators not equal of 0, 1, or 2, or negative share sizes or prices were excluded from calculations by the algorithm.

Following techniques from \cite{intraday1,intraday2,intraday3}, the measured variable is the traded value per unit time where the traded value$V_i(n)$ for a stock $i$ in transaction $n$ is defined as

\begin{equation}
V_i(n) = p_i(n)S_i(n)
\end{equation}

where $p_i(n)$ is the price the trade is executed at and $S_i(n)$ is the share size of the trade. The traded value per unit time $\Delta t$ is defined as
\begin{equation}
f_i^{\Delta t}(t) = \sum_{n,t_i(n)\in[t,t+\Delta t]}V_i(n)
\end{equation}

The trade data had a resolution of 1 second and trades were aggregated into 1 second buckets for analysis of the time series. Each trading day was analyzed independently and data was not combined across days. 
\begin{table*}[!tc] \vspace{1.5ex}
\centering
\begin{tabular}{|c|c|}
\hline
Quote Symbol&Stock Name\\
\hline
BAC&Bank of America \\
C&Citicorp\\
PG&Proctor \& Gamble\\
GE&General Electric\\
ITT&ITT Industries \\
CHD&Church \& Dwight\\
PFE&Pfizer\\
MSFT&Microsoft\\
INTC&Intel\\
CSCO&Cisco\\
GENZ&Genzyme\\
GILD&Gilead Sciences\\
AAPL&Apple Computer Inc.\\
NWS&News Corp.\\
\hline
 \end{tabular}
\caption{Quote symbols and stocks discussed in this paper.}
\label{stocklist}
\end{table*}
The first step was a brief analysis to investigate the presence or absence of HFT via average share sizes over the measured period. For all selected stocks, there was a marked decrease in the average number of shares per trade shown collectively in figure \ref{sharesize}. This decrease in trade size was used as prima facie evidence of the influence of HFT on the selected stocks. Next to be addressed was whether the short-term correlation structure of intraday stock trading has been significantly affected by HFT and related strategies. All previous papers, and data pulled from NYSE TAQ on stocks in 1993 by the author, indicate that short-term correlations were traditionally largely absent from the trading patterns, consistently showing an average $H \approx 0.5$.

In order to fully investigate the structure of intraday trading over several time scales at many orders of magnitude, a discrete wavelet transform was used to create a logscale diagram to observe the behavior of the second moment of the detailed wavelet coefficients across multiple octaves representing different time scales. A brief overview of the wavelet analysis is given in the appendix and the logscale diagram in described in the next section.

\section{Discrete wavelet transforms and logscale diagrams}

Wavelets were chosen to understand the data for a variety of reasons. First, common methods used to measure the Hurst exponent such as rescaled range analysis (R/S statistic) or aggregated variance analysis are built around an assumption of stationarity in the data. This assumption is not valid for intraday stock time series since there is a well-known and marked non-stationarity including peaks of activity at the beginning and end of the day and a dip during lunch time. R/S scaling has also been shown to introduce distortions in the measurement of the Hurst exponent in financial time series \cite{RScriticize}. Second, some common methods to analyze the correlation structure of nonstationary data such as detrended fluctuation analysis (DFA) have been found to systematically introduce errors in measurements of the Hurst exponent except in the case of simple monotonic trends \cite{DFAcriticize}. 

The mathematical preliminaries of wavelets and the logscale diagrams will be skipped here but shown in the appendix. The analysis here follows the detailed methodology of Abry, Flandrin, Taqqu, \& Veitch \cite{hurstguide1,hurstguide2}. This analysis uses a discrete wavelet transformation (DWT) over 14 octaves using the Haar wavelet. The detailed coefficients $d_X(j,k)$ for each octave, $j$, and time slice $2^jk$ were used to calculate the second moment of the coefficients for each octave following the equation

\begin{equation}
S_2(j) = \frac{1}{n_j}\sum_k |d_X(j,k)|^2
\end{equation}

The logscale diagram is then created by plotting $y_j=\log_2 S_j(j)$ vs. $j$ using a logarithm of base 2. The subsequent step is to analyze the logscale diagrams for each stock in a variety of time periods to identify characteristic trends. As stated in the appendix, the presence of self-similar behavior representing $H>0.5$ is indicated by a positive linear slope across multiple octaves. This scale-invariant (here scale is over different time scales) behavior is a key feature of self-similarity. As a guide, a minimum trend of three points is necessary to unambiguously characterize self-similar behavior across multiple time scales \cite{hurstguide1,hurstguide2}.

\begin{figure}
\centering
\includegraphics[height=2.75in,width=2.75in]{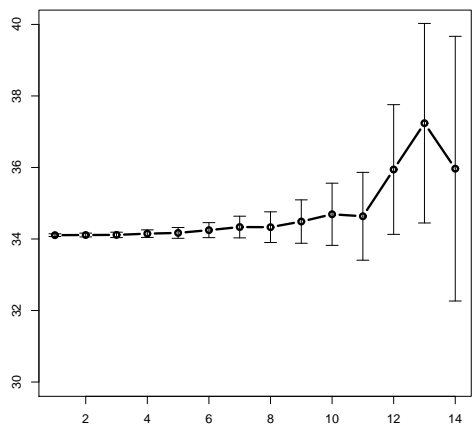}
\caption{Typical logscale diagram. Average of octave values for PG over 2005. Error bars represent the 95\% confidence intervals for each octave.}
\label{logscale}

\end{figure}

The logscale diagrams for the intraday time fluctuations of total value all seem to show common features as seen in figure \ref{logscale}. One can distinguish biscaling behavior over two different time scales. On the high frequency end, octaves $j\in [1,10]$, there is a relatively flat trend, especially in the earliest years. Given that the Hurst exponent is related to the slope $\alpha$ of the logscale diagram by

\begin{equation}
H = \frac{\alpha+1}{2}
\label{hurstalpha}
\end{equation}

a flat slope indicates $H=0.5$ and normal Brownian motion. At $j\in[10,11]$ there is usually a transition, often involving a decrease in $y_j$ and then another, steeper, positive trend over the next couple of octaves though there is sometimes a dip at $j=14$, possibly due to boundary effects of the signal size. There are two key features of this second, low-frequency region of greater than $2^{11}$ seconds (about 34 minutes). First, it seems to represent correlations that appear over substantial periods of time in the trading day from many minutes to hours. Second, it is present over all years with data including all the way back to 1993. This seems to indicate that this is a long-extant feature of the intraday stock trading dynamics and not a new manifestation of behavior. With greater resolution data, we would like be able to robustly calculate a Hurst exponent $H>0.5$ for this region over all years. This finding does not contradict earlier work that showed an exponent of $H\approx 0.5$ for traded value during a trading day since the previous research only looked at intraday time scales under 60 minutes which would not resolve these features \cite{intraday1,intraday2,intraday3}. The last thirty seconds of trading were excluded from the NASDAQ time series since on some days huge spikes in share trading and the ending of trading often skewed the wavelet coefficients and distorted logscale diagrams despite trading throughout the day following the traditional pattern.

\subsection{Logscale diagrams and Hurst exponent estimation}

Therefore, the focus of any search for the effects of recent HFT should be focused on the high frequency region from octaves 1-10. A simple unbiased method of calculating the slope of the logscale diagram in this region would be least-squares regression, however, it is more advisable to use the method of \cite{hurstguide1, hurstguide2} where a weighted regression that weights each $S_2(j)$ by its variances $\sigma_j^2$ to calculate the minimum variance unbiased estimator correcting for heteroscedasticity.

The estimator $\hat{\alpha}$ for $\alpha$ over an octave range $j\in[j_1,j_2]$ is given by

\begin{equation}
\hat{\alpha}=\frac{\sum_j y_j(Sj - S_j)/\sigma_j^2}{SS_{jj} - S_j^2} \equiv \sum_j w_jy_j
\label{estimator}
\end{equation}

where the variance $\sigma_j^2$ is defined as
\begin{equation}
\sigma_j^2 = \frac{\zeta(2,n_j/2)}{\ln^2 2}
\end{equation}

Where $\zeta(z,v)$ is a generalized Riemann Zeta function and $n_j$ is the number of detailed coefficients in the octave $j$. Note that the variance depends only on the number of coefficients and not the details of the data.

The additional variables are defined $S=\sum_{[j_1,j_2]}(1/\sigma_j^2)$, $S_j=\sum_{[j_1,j_2]}(j/\sigma_j^2)$ and $S_{jj}=\sum_{[j_1,j_2]}(j^2/\sigma_j^2)$. Note the $S$ variables must be calculated before the summation in the full estimator equation. Additionally, the variance of $\hat{\alpha}$ is given by

\begin{equation}
Var(\hat{\alpha}) = \sum_j \sigma_j^2 w_j^2
\end{equation}

The distribution of $\hat{\alpha}$, can be considered approximately Gaussian and using two standard deviations for 95\% confidence intervals, the 95\% confidence interval of $\hat{\alpha}$ $\pm 0.02$ and given equation \ref{hurstalpha}, the 95\% confidence interval of calculated daily $H$ is $\pm  0.01$.

\begin{figure}
\centering
\begin{tabular}{cc}
\includegraphics[height=2.75in,width=2.75in]{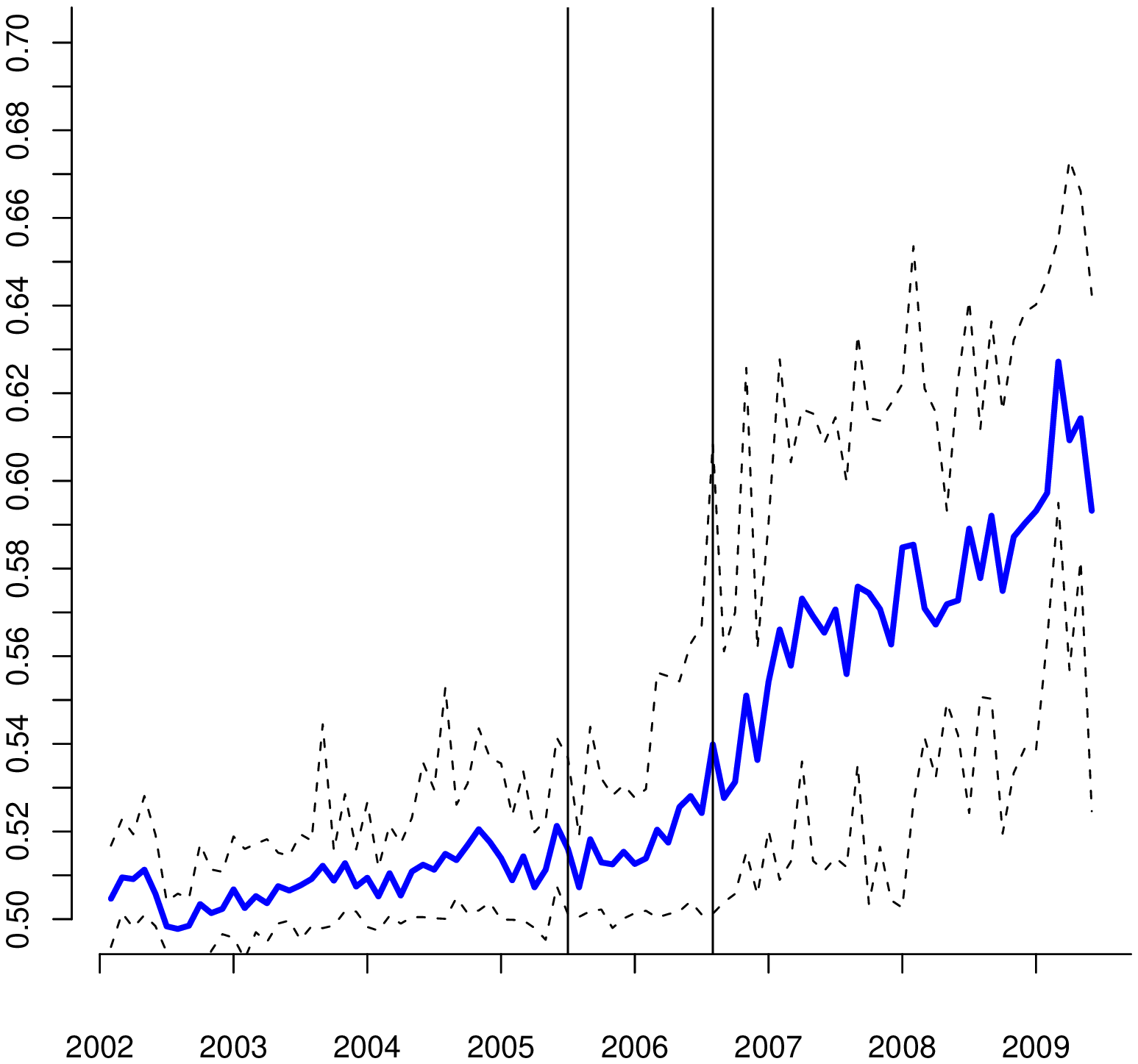}& 
\includegraphics[height=2.75in,width=2.75in]{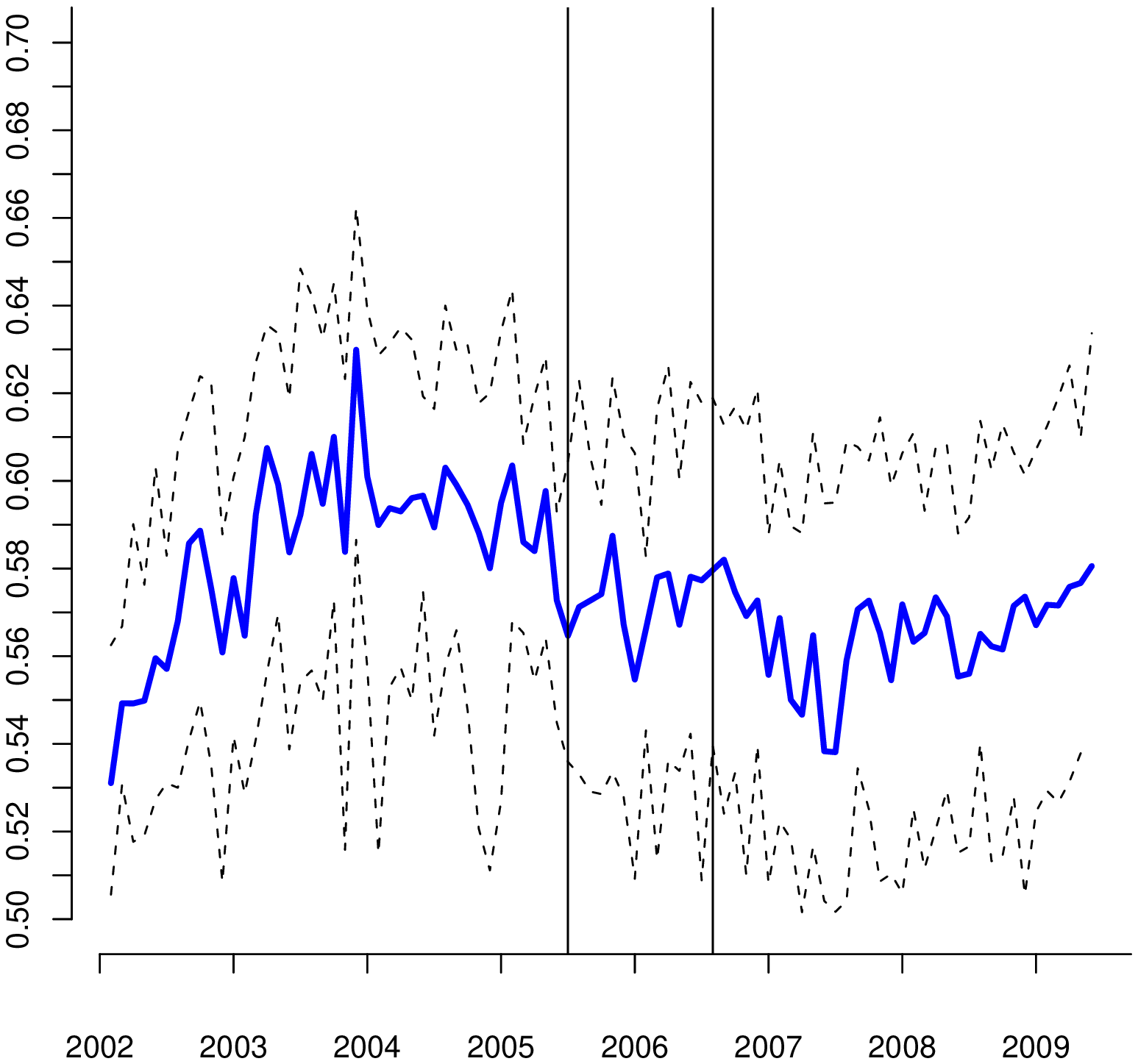}\\
\includegraphics[height=2.75in,width=2.75in]{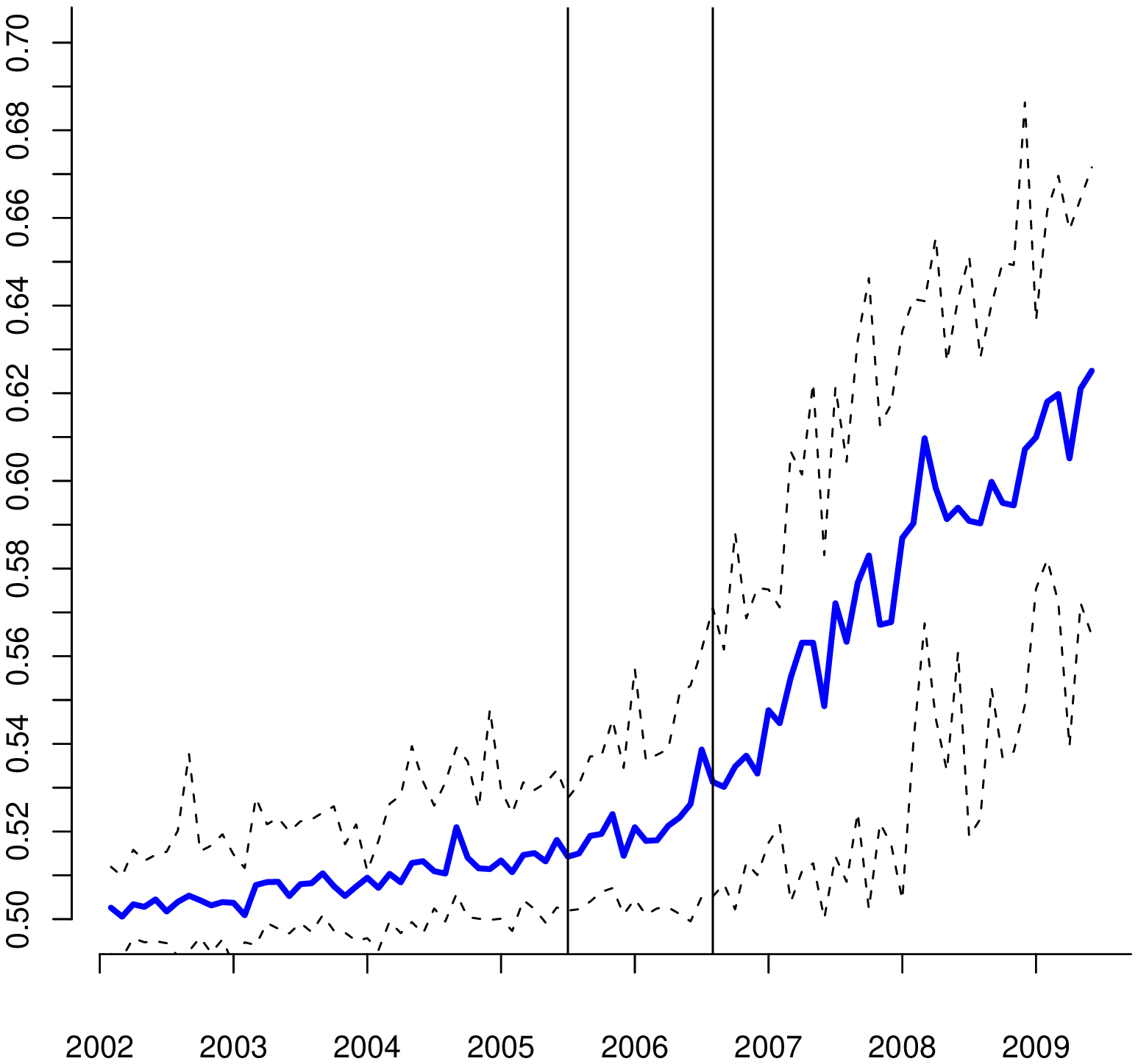}& 
\includegraphics[height=2.75in,width=2.75in]{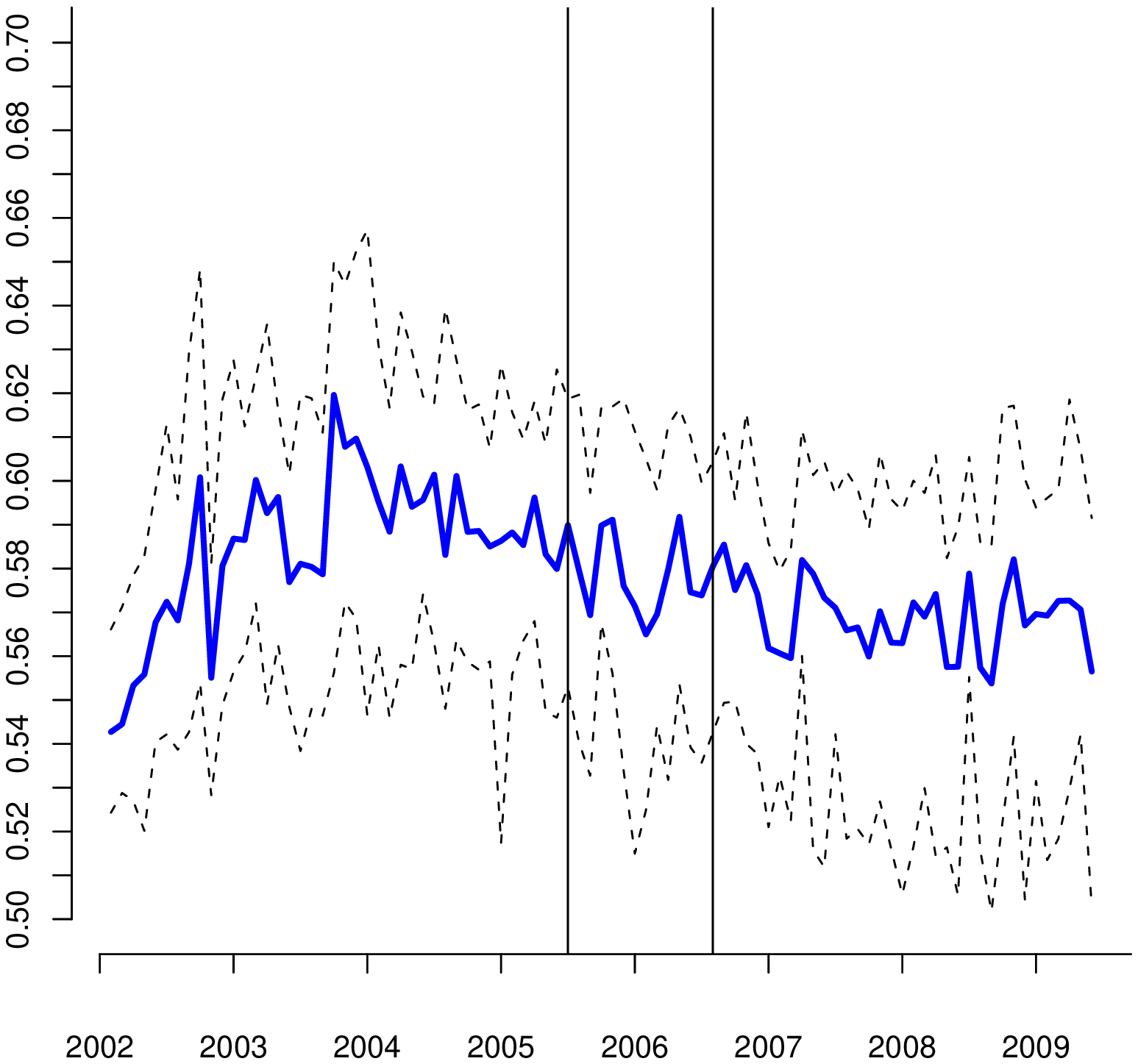}\\
\includegraphics[height=2.75in,width=2.75in]{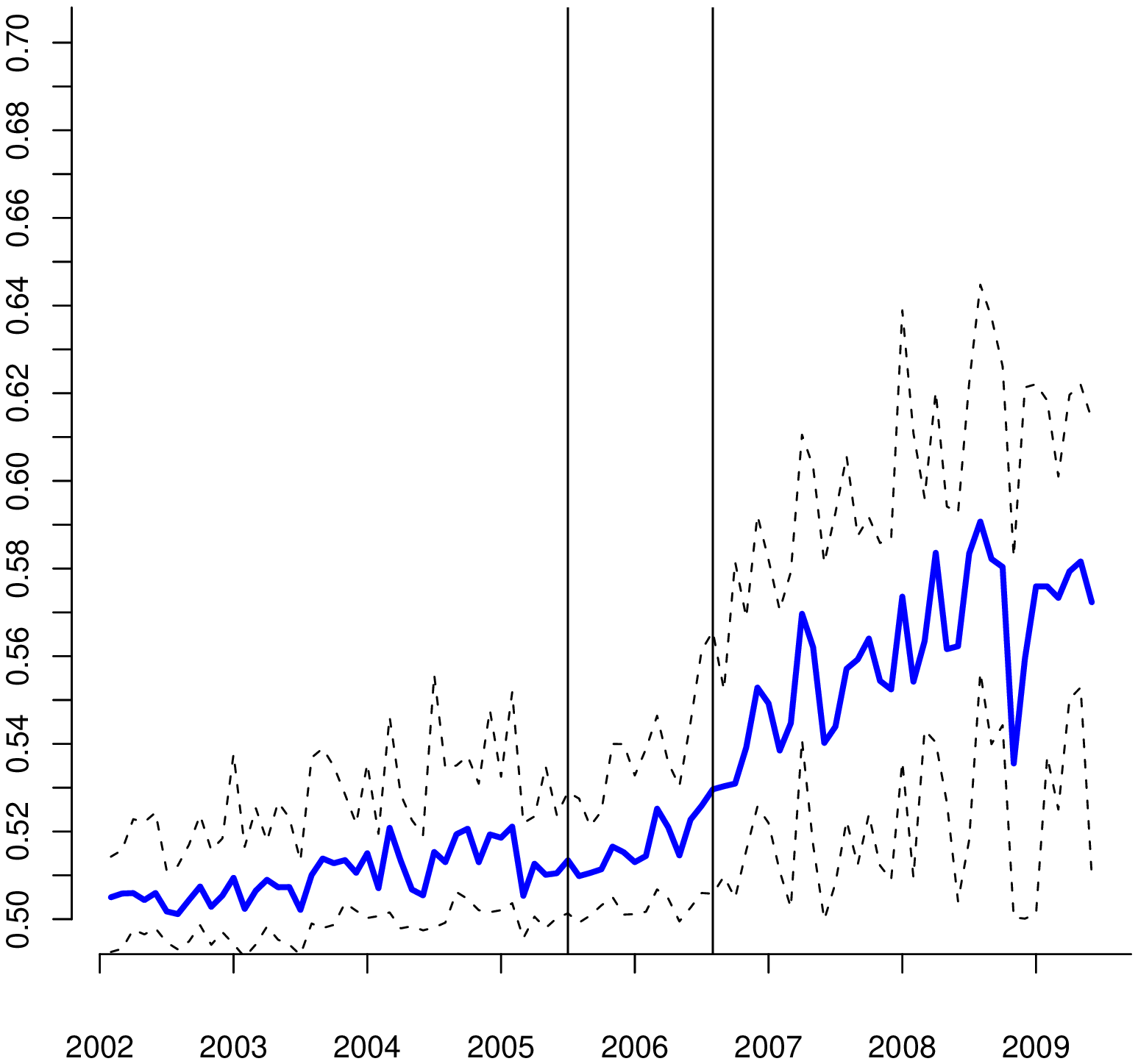}& 
\includegraphics[height=2.75in,width=2.75in]{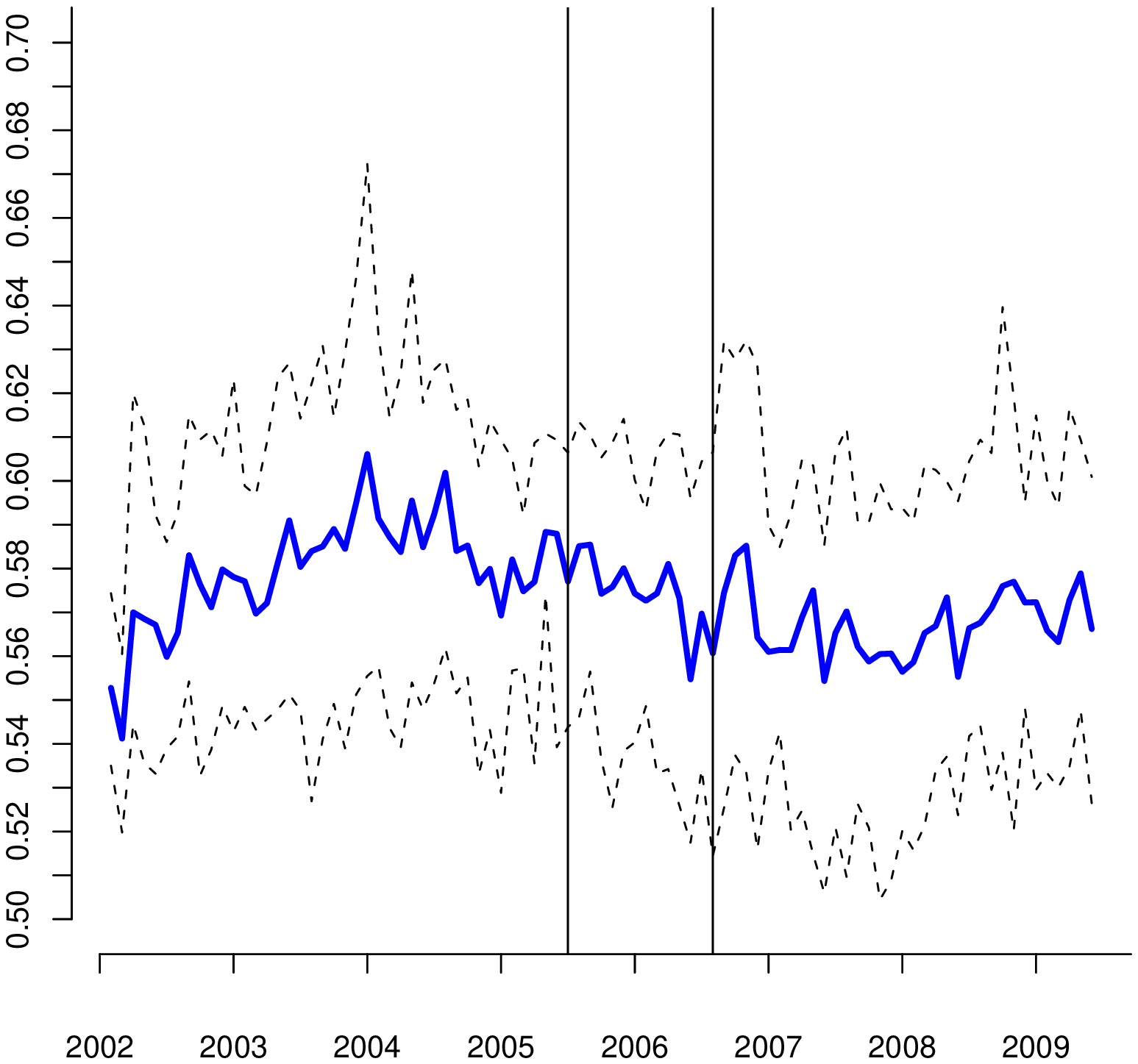}\\
\end{tabular}
\caption{The average Hurst exponent by month over the time period of 2002-May 2009 as calculated using the wavelet coefficient method. The blue line represents the average value while the dotted lines above and below represent the 95\% confidence interval of values during each month. The two black vertical lines represent first the promulgation of Reg NMS by the SEC in June 2005 and the first implementation compliance date of June 2006. Stocks in each row are paired by NYSE/NASDAQ: Bank of America (BAC)/Microsoft (MSFT), Citigroup (C)/Intel (INTC), Proctor \& Gamble (PG)/Cisco Systems (CSCO)).}
\label{stockhursts1}
\end{figure}

\begin{figure}
\centering
\begin{tabular}{cc}
\includegraphics[height=2.75in,width=2.75in]{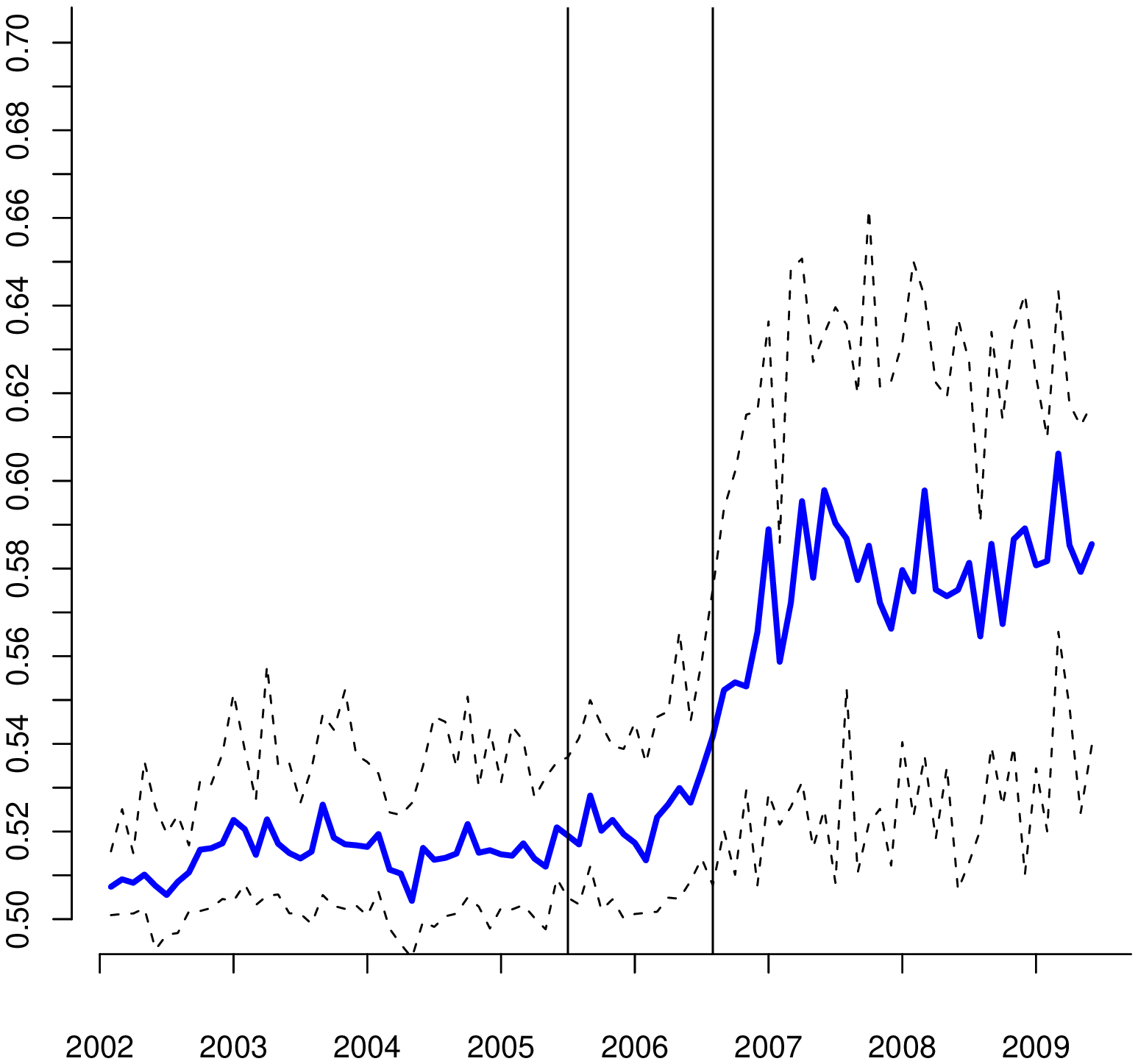}& 
\includegraphics[height=2.75in,width=2.75in]{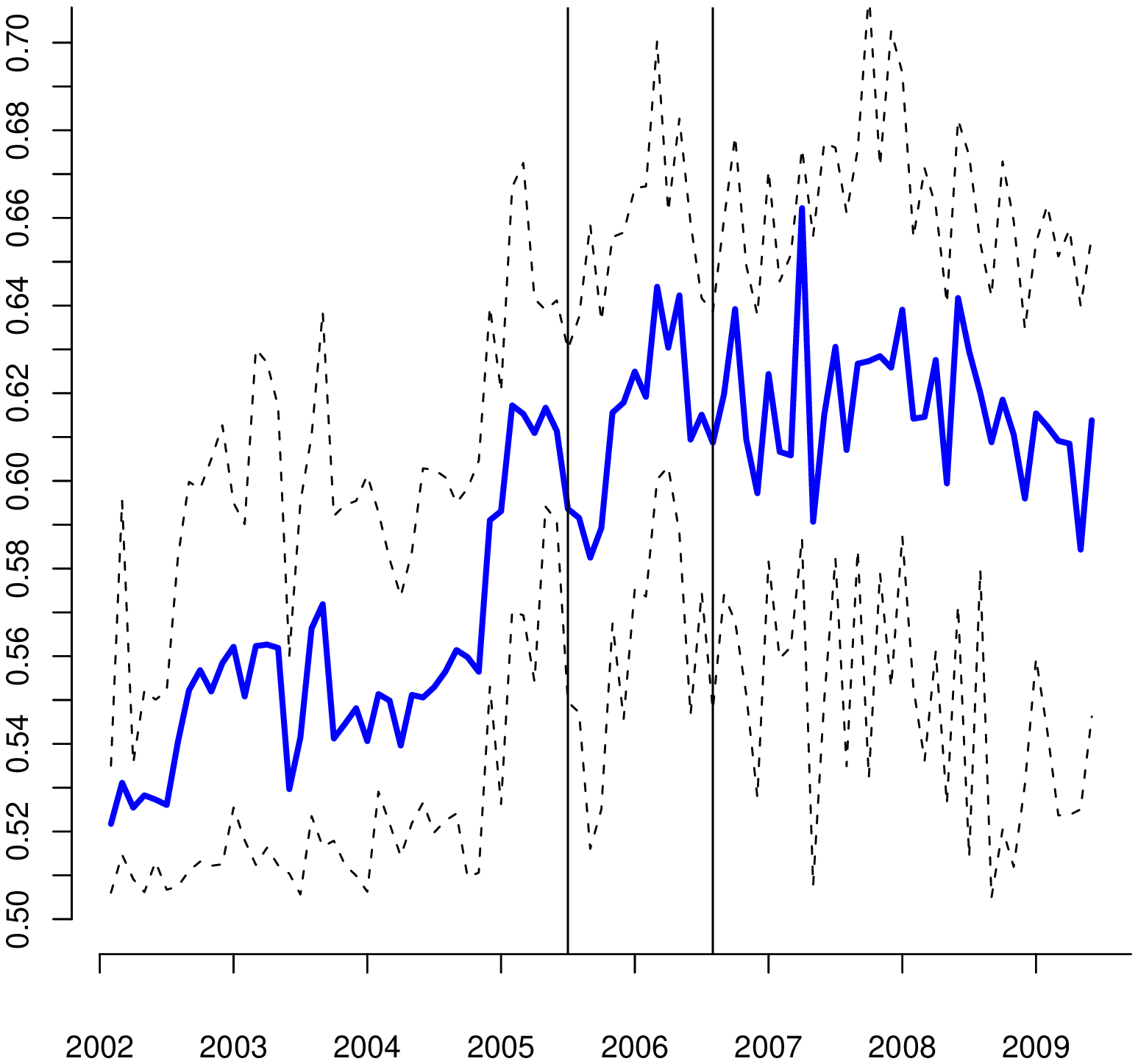}\\
\includegraphics[height=2.75in,width=2.75in]{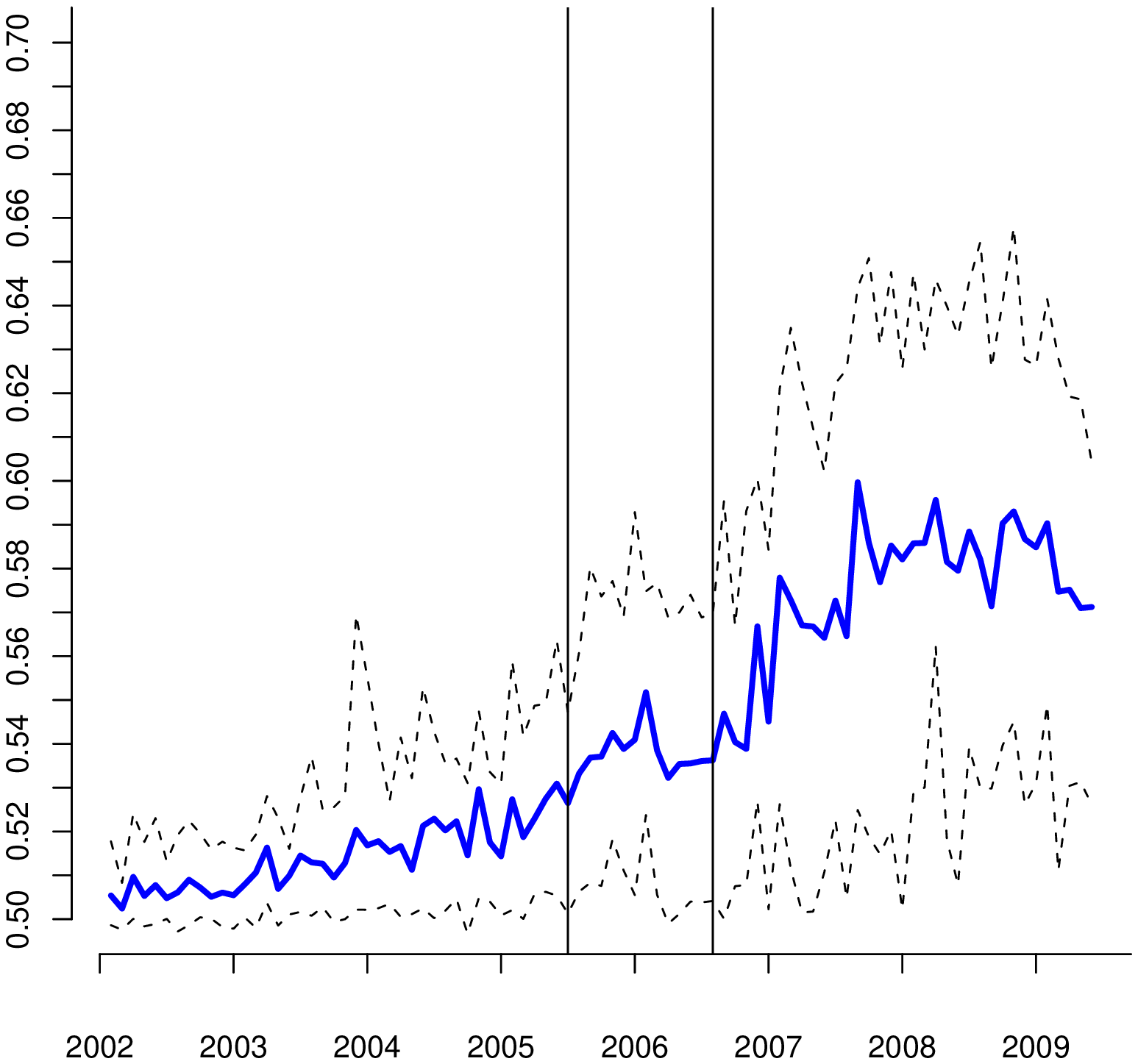}& 
\includegraphics[height=2.75in,width=2.75in]{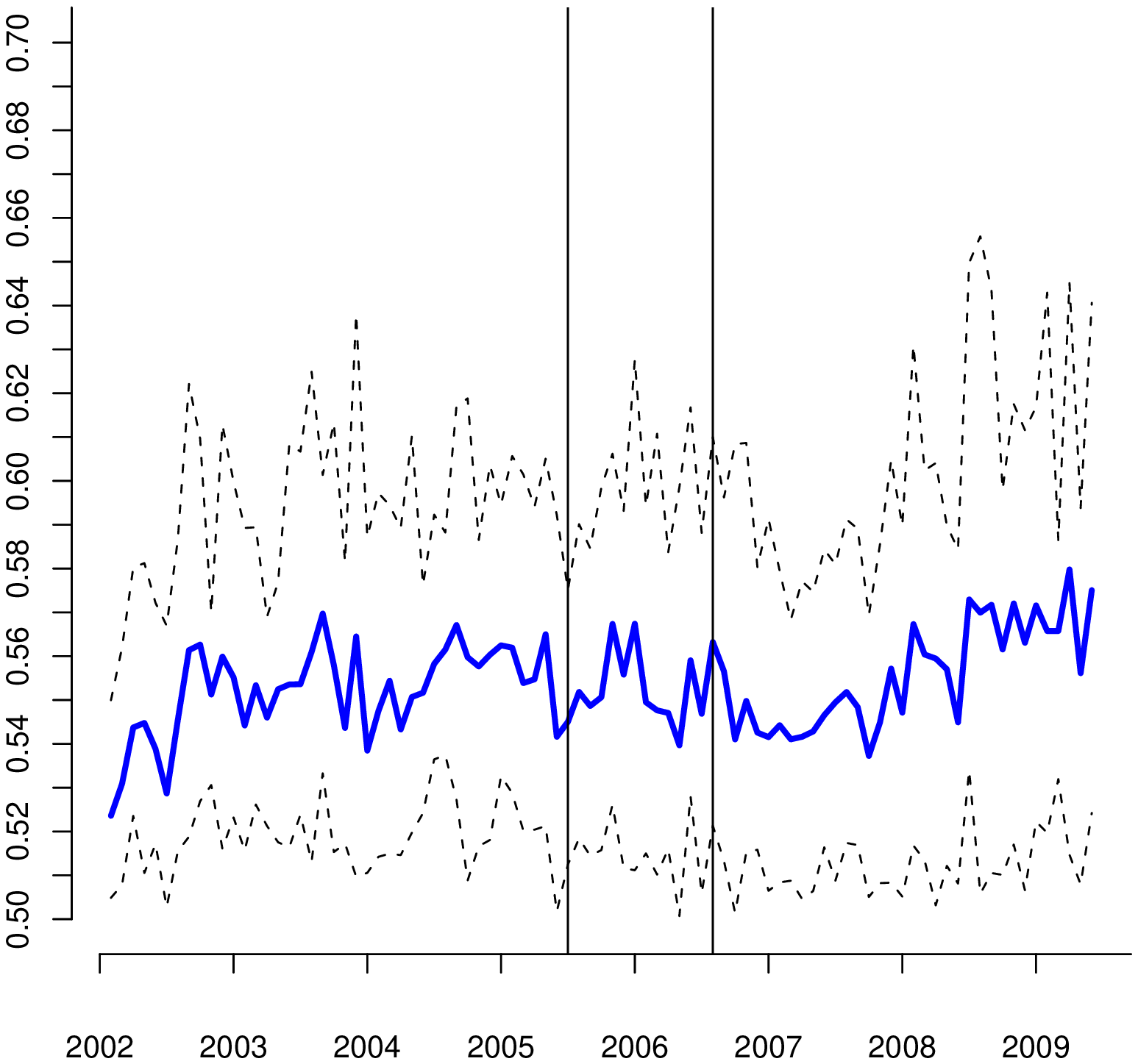}\\
\includegraphics[height=2.75in,width=2.75in]{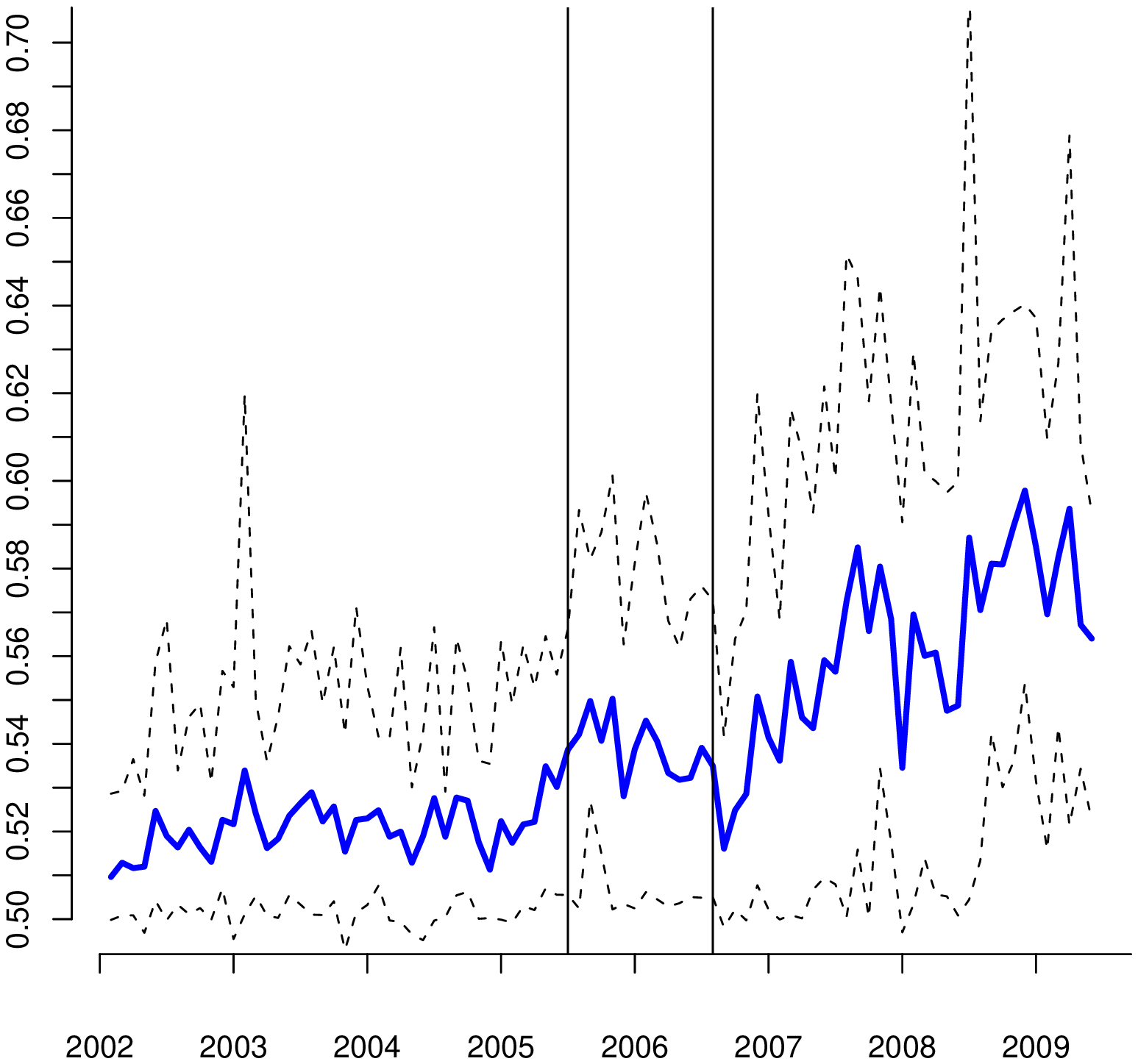}& 
\includegraphics[height=2.75in,width=2.75in]{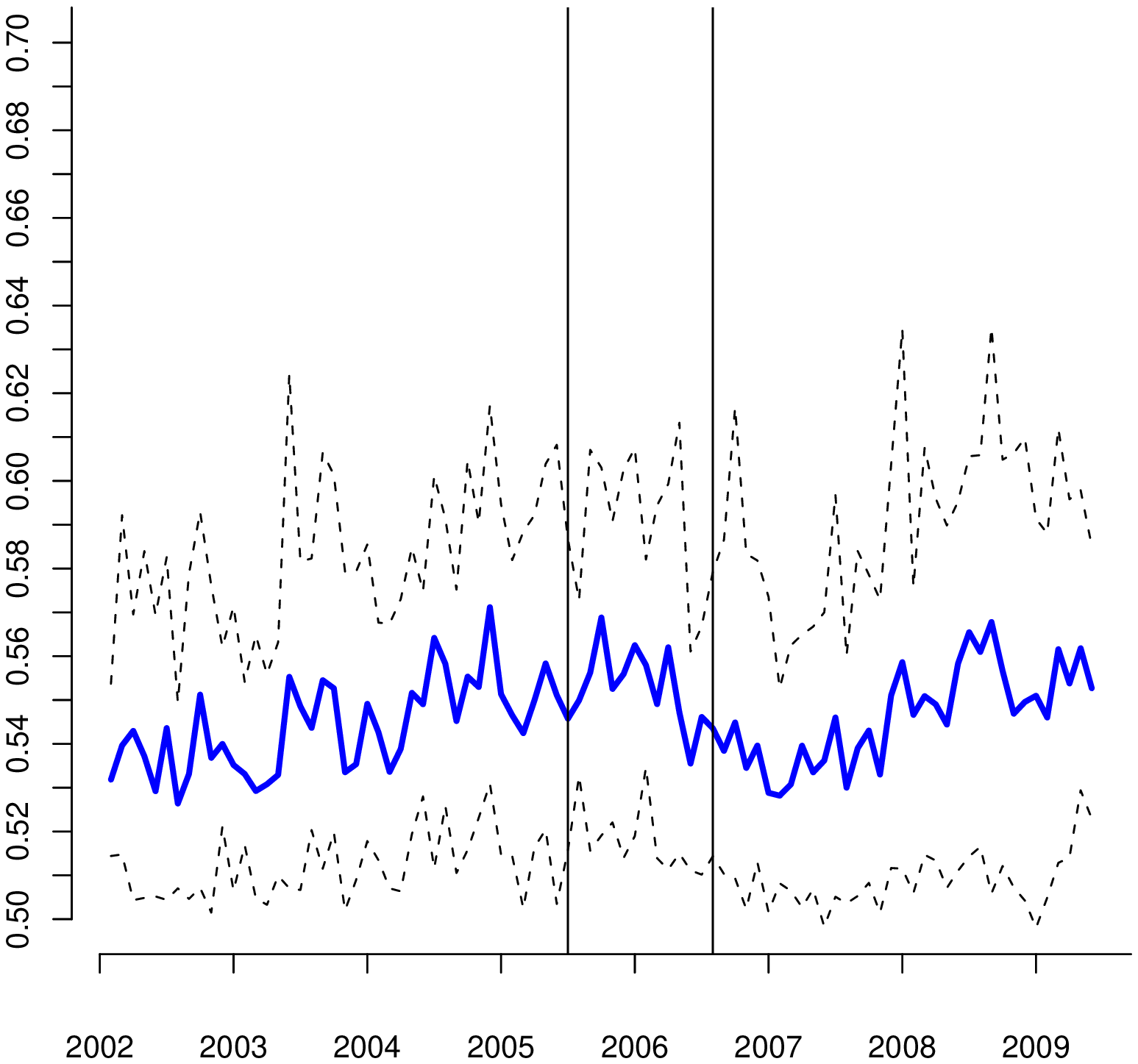}\\
\end{tabular}
\caption{The average Hurst exponent by month over the time period of 2002-May 2009 as calculated using the wavelet coefficient method. The blue line represents the average value while the dotted lines above and below represent the 95\% confidence interval of values during each month. The two black vertical lines represent first the promulgation of Reg NMS by the SEC in June 2005 and the first implementation compliance date of June 2006. Stocks in each row are paired by NYSE/NASDAQ: General Electric (GE)/Apple Computer (AAPL), ITT Industries (ITT)/Genzyme (GENZ), Church \& Dwight (CHD)/Gilead Sciences (GILD).}
\label{stockhursts2}
\end{figure}

\begin{figure}
\centering
\begin{tabular}{cc}
\includegraphics[height=2.75in,width=2.75in]{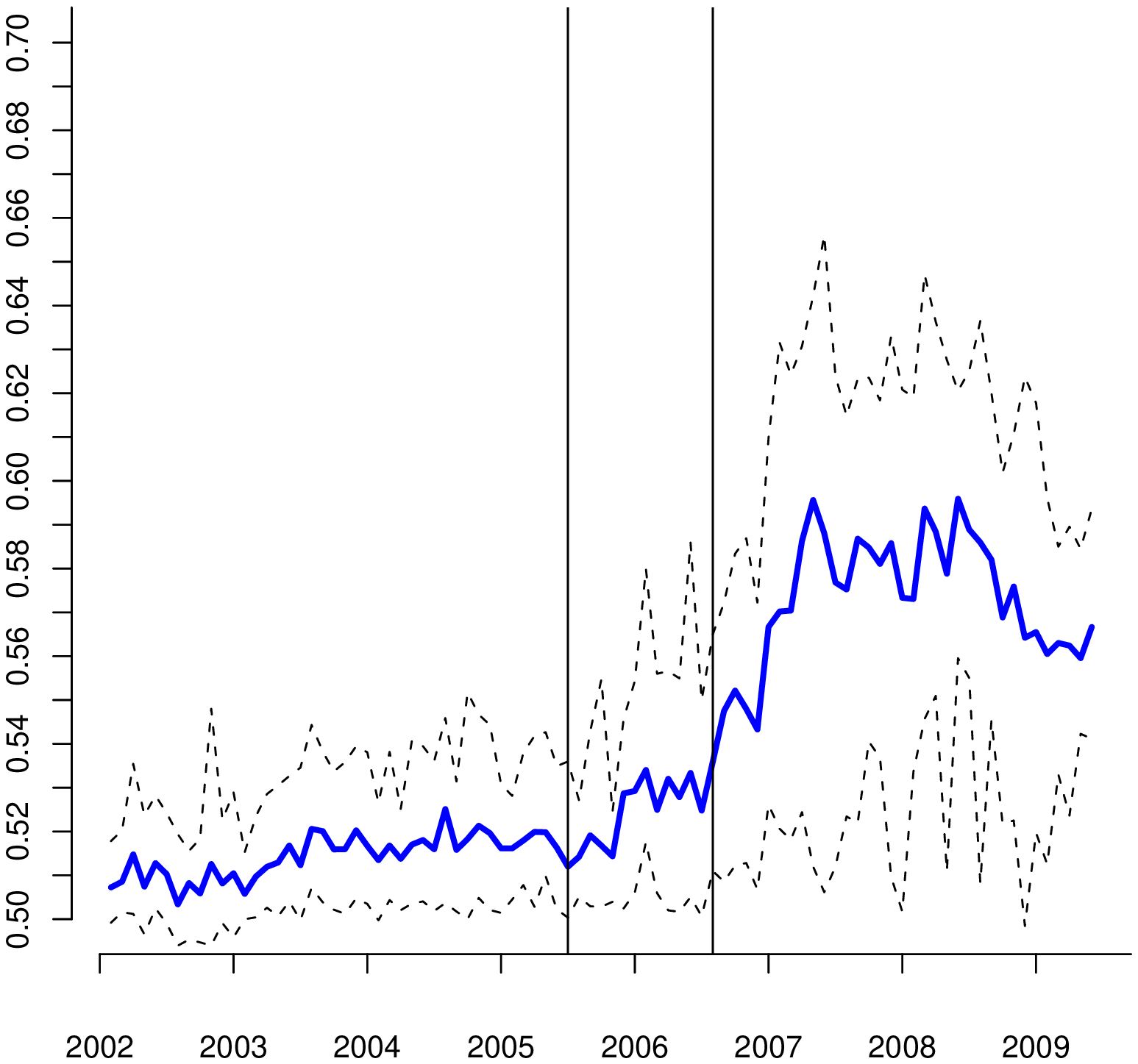}& 
\includegraphics[height=2.75in,width=2.75in]{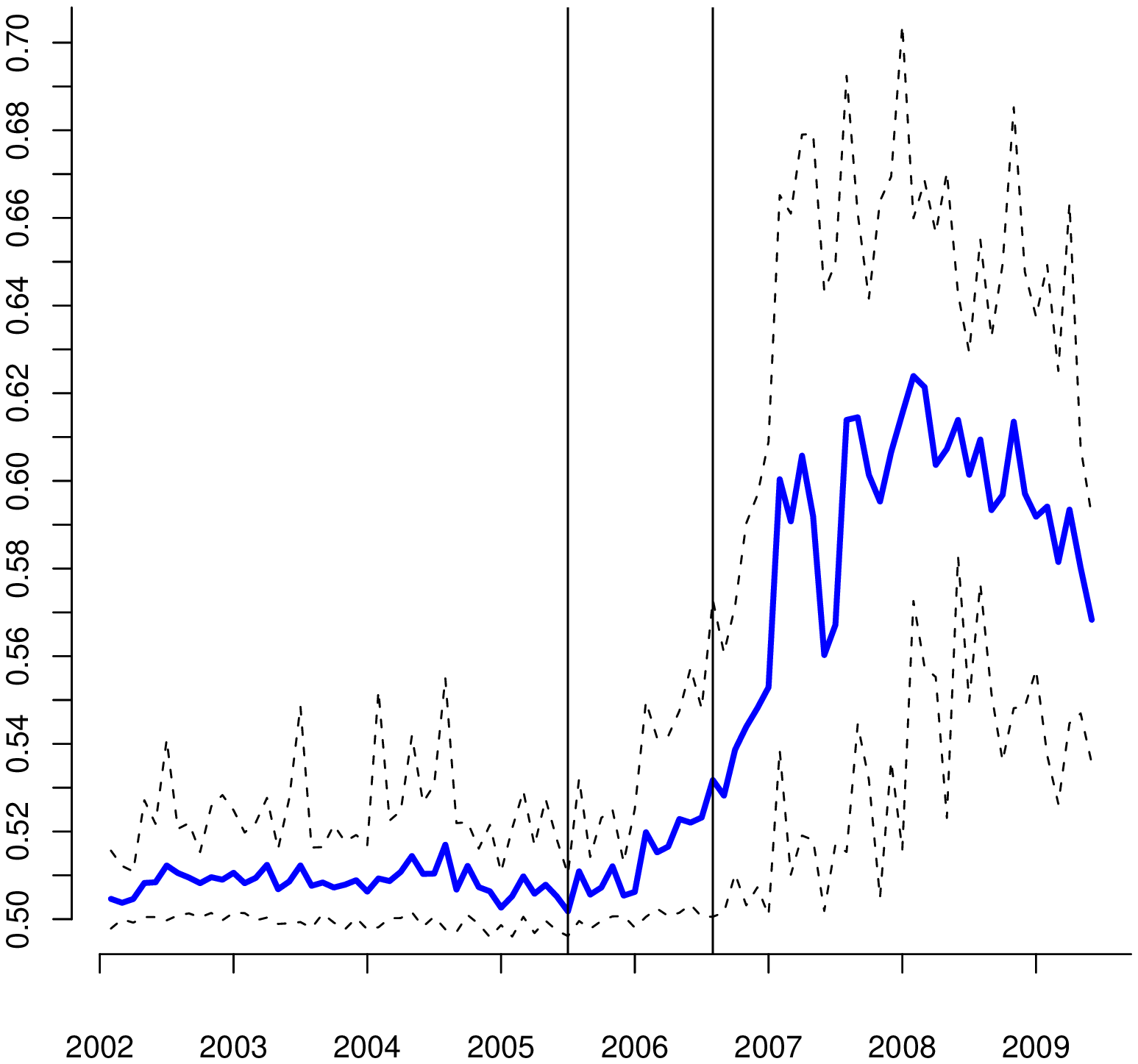}\\
\end{tabular}
\caption{The average Hurst exponent by month over the time period of 2002-May 2009 as calculated using the wavelet coefficient method. The blue line represents the average value while the dotted lines above and below represent the 95\% confidence interval of values during each month. The two black vertical lines represent first the promulgation of Reg NMS by the SEC in June 2005 and the first implementation compliance date of June 2006. Stocks in each row are paired by NYSE/NASDAQ: Pfizer (PFE)/News Corp. (NWS).}
\label{stockhursts3}
\end{figure}

\section{Results of wavelet analysis}

Calculations of $H$ according to the above methodology were done for each stock on each day over the time periods of the data. In order to more easily visualize trends, $H$ was averaged over each month and plotted over time from January 2002 to May 2009. The following pages show the figures for monthly averages of the trends for each stock over time with 95\% confidence intervals, given the range of measured $H$ over the month, given by dashed lines. The two vertical solid lines give the passing of the Reg NMS changes and their first implementation deadline respectively.

The first clear feature is a trend across all stocks for the time periods shown. For the NYSE stocks, the Hurst exponents increase from 2002 onward but by late 2005 barely break the average of $H=0.55$. Therefore, during this time (and before), short term trading fluctuations do not appreciably depart from an approximation of Gaussian white noise. However, once Reg NMS is implemented the structure of the trading noise begins to change rapidly increasing to 0.6 and beyond in a couple of years. This is a new behavior in the high-frequency spectrum of trading data that indicates increasingly correlated trading activity over increasingly shorter timescales over the last several years. Correlations previously only seen across hours or days in trading time series are increasingly showing up in the timescales of seconds or minutes. 

A more complicated picture is shown in the NASDAQ data. The Hurst exponent of NASDAQ stocks started rising much earlier, from 2002 or earlier. By the time Reg NMS was passed, most NASDAQ stocks already had an $H$ which many stocks on the NYSE would not reach until 2009. Some NASDAQ stocks, such as AAPL or CSCO, did have spikes shortly after the new rules went into effect but soon returned to normal behavior. In fact, data from Figure 10 shows this spike for AAPL was likely not due to HFT as it was most pronounced amongst larger sized trades. It is probable that since NASDAQ was one of the first exchanges to embrace electronic trading and experience HFT via ECNs, the rules officially unchaining HFT for other exchanges from 2005 was close to a non-event. Possibly, because HFT was experienced earlier, there is not as dramatic a rise from the time of Reg NMS in $H$ for NASDAQ. The one marked change from the implementation time of Reg NMS is the emergence of massive share trades at the final seconds of trading days, sometimes varying over many orders of magnitude by day, that required the time series to remove to get clear data.

In order to more clearly understand the source of the increased self-similarity in the trading noise, the data was analyzed again in buckets corresponding to trades of a certain size. In all cases it was found that the $H$ of trades where the average shares/trade was greater than about 1500 shares had an $H\approx 0.5$ for all time periods while $H > 0.5$ for trades with an average of less than 1500 shares per trade. The results are shown for several stocks in the first columns of figures 9 and 10. It shows that over all time periods, trades with smaller share sizes show a higher $H$ than trades with larger share sizes. The relative difference does not change appreciably over time. Trades with more than 1500 shares per trade are relatively moribund and do not usually depart from $H\approx 0.5$. However, as the trade sizes decrease, especially to 1500 shares or below, there is a high level of correlation in the trading activity. The relative activity of each of these trade sizes stays relatively proportional over time.

Therefore, stock trading noise as measured here can be determined to be a superposition of two types of noise. For large share trades, the noise is approximately Gaussian white noise, while for small share trades the noise has a more fractal character. So if the relative $H$ for share trade sizes stays relatively constant over time why does the overall $H$ increase? The answer is illustrated in the second column of figures 9 and 10 which shows how over time, and especially since the boom of widespread HFT, the proportion of all trades by the small share trades has increased markedly. Therefore, we can determine the increasing $H$ of overall stock trading noise is generated by a larger relative proportion of trading by small share trades generated primarily through HFT. The dynamics of HFT and non-HFT trading have not altered much, but their relative magnitude has changed dramatically.

\begin{figure}
\centering
\begin{tabular}{cc}
\includegraphics[height=2.75in,width=2.75in]{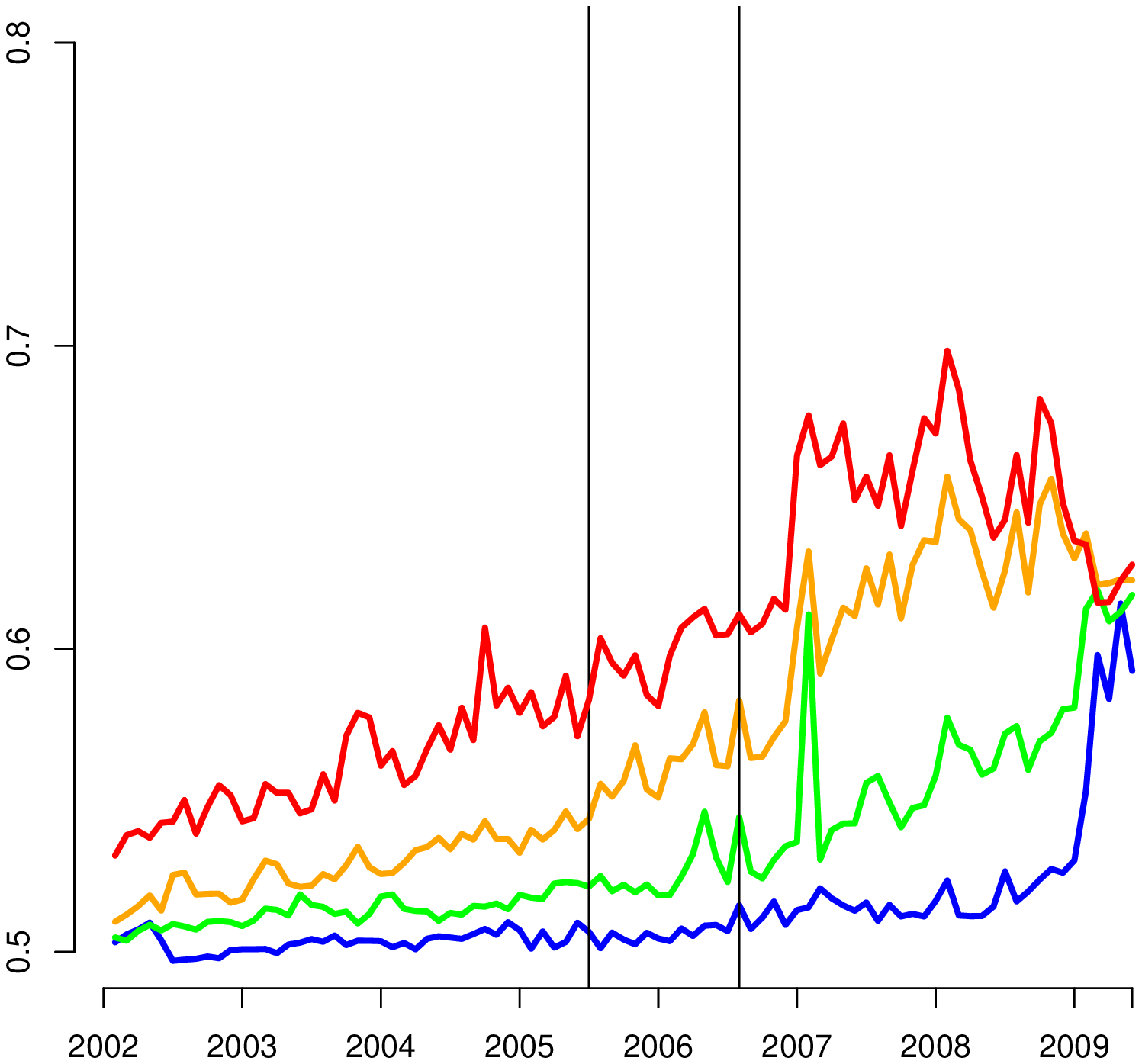}& 
\includegraphics[height=2.75in,width=2.75in]{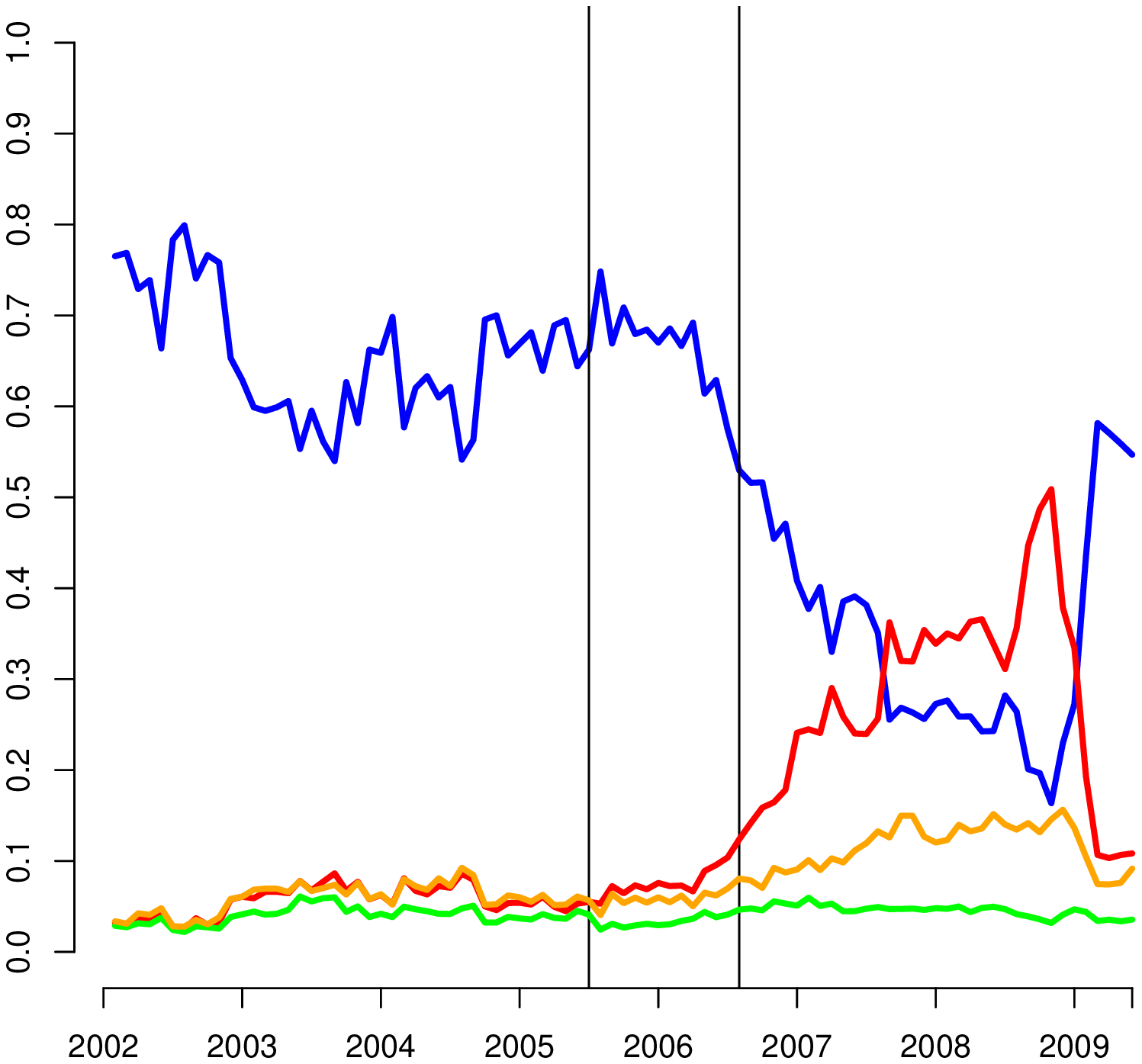}\\
\includegraphics[height=2.75in,width=2.75in]{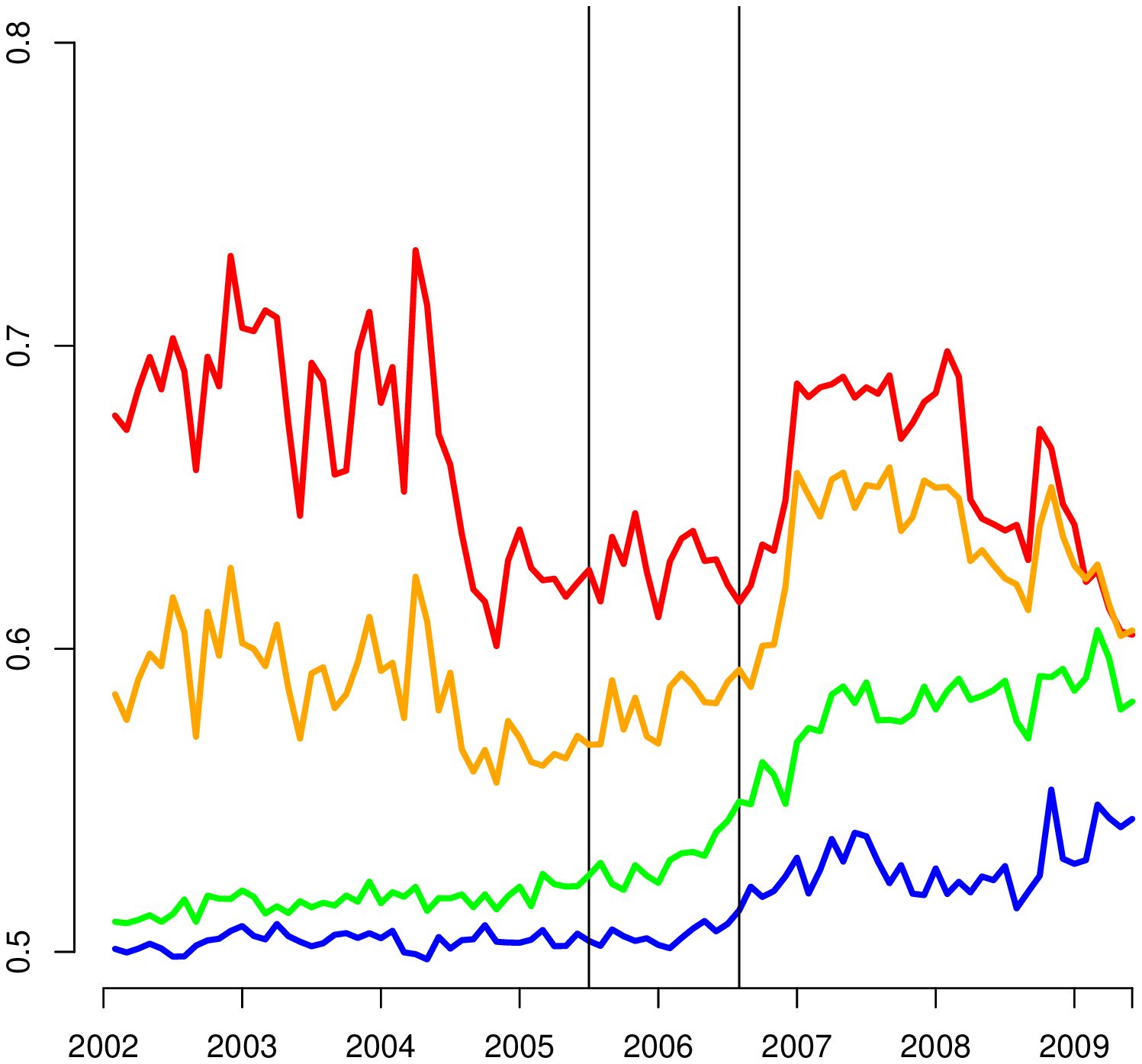}& 
\includegraphics[height=2.75in,width=2.75in]{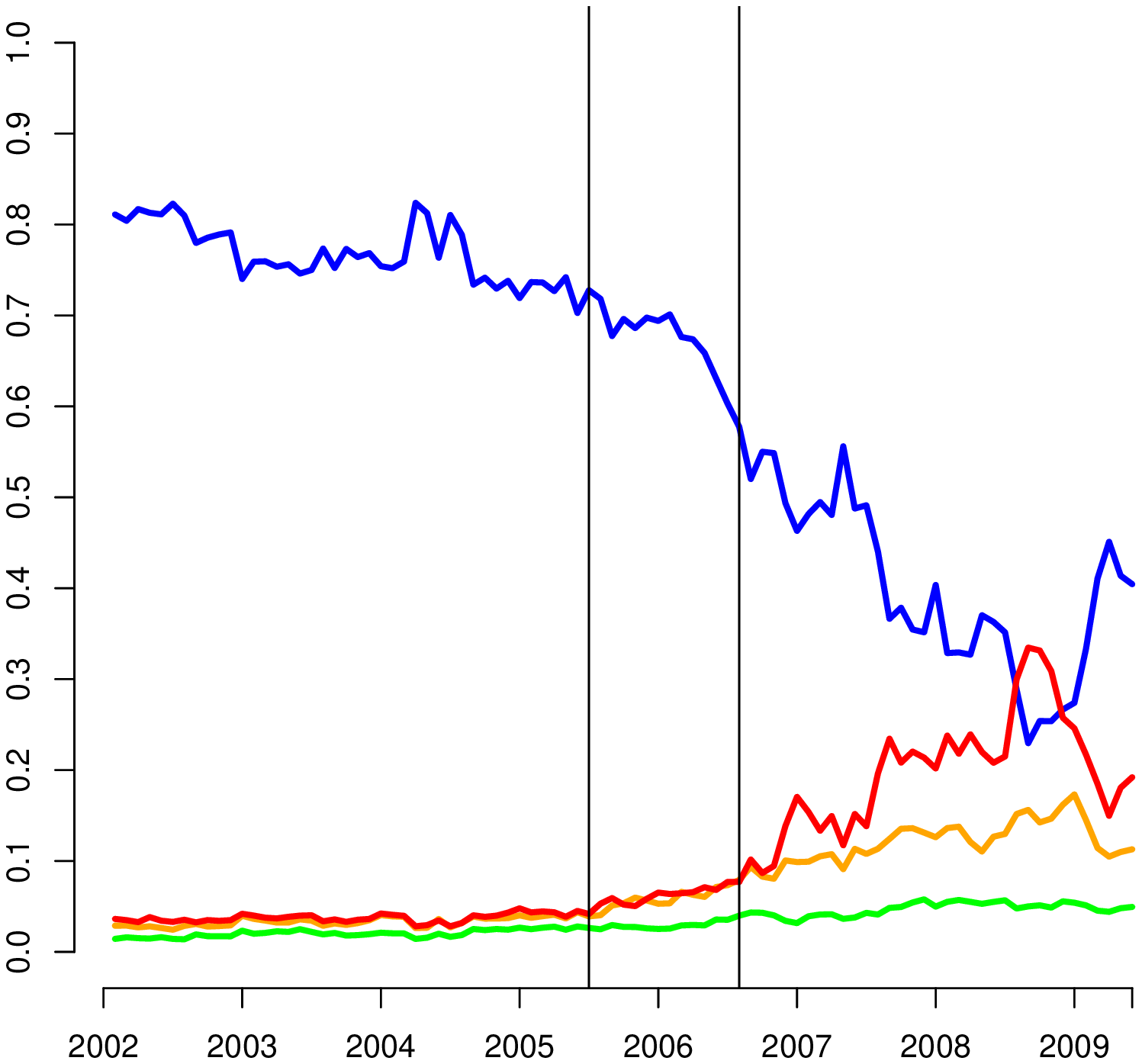}\\
\end{tabular}
\label{3dhurst}
\caption{Each row of images shows the average Hurst exponent
 by month for a range of trade sizes in the first image. The second image is the relative proportion of all trades by that trade size range. Row by row are, Bank of America (BAC) and General Electric (GE). The color lines indicate the range of shares per trade analyzed: blue is 1500+ shares per trade, green is 750-1000 shares per trade, orange is 250-500 shares per trade, and red is less than 250 shares per trade.}
\end{figure}
\pagebreak 
\begin{figure}
\centering
\begin{tabular}{cc}
\includegraphics[height=2.75in,width=2.75in]{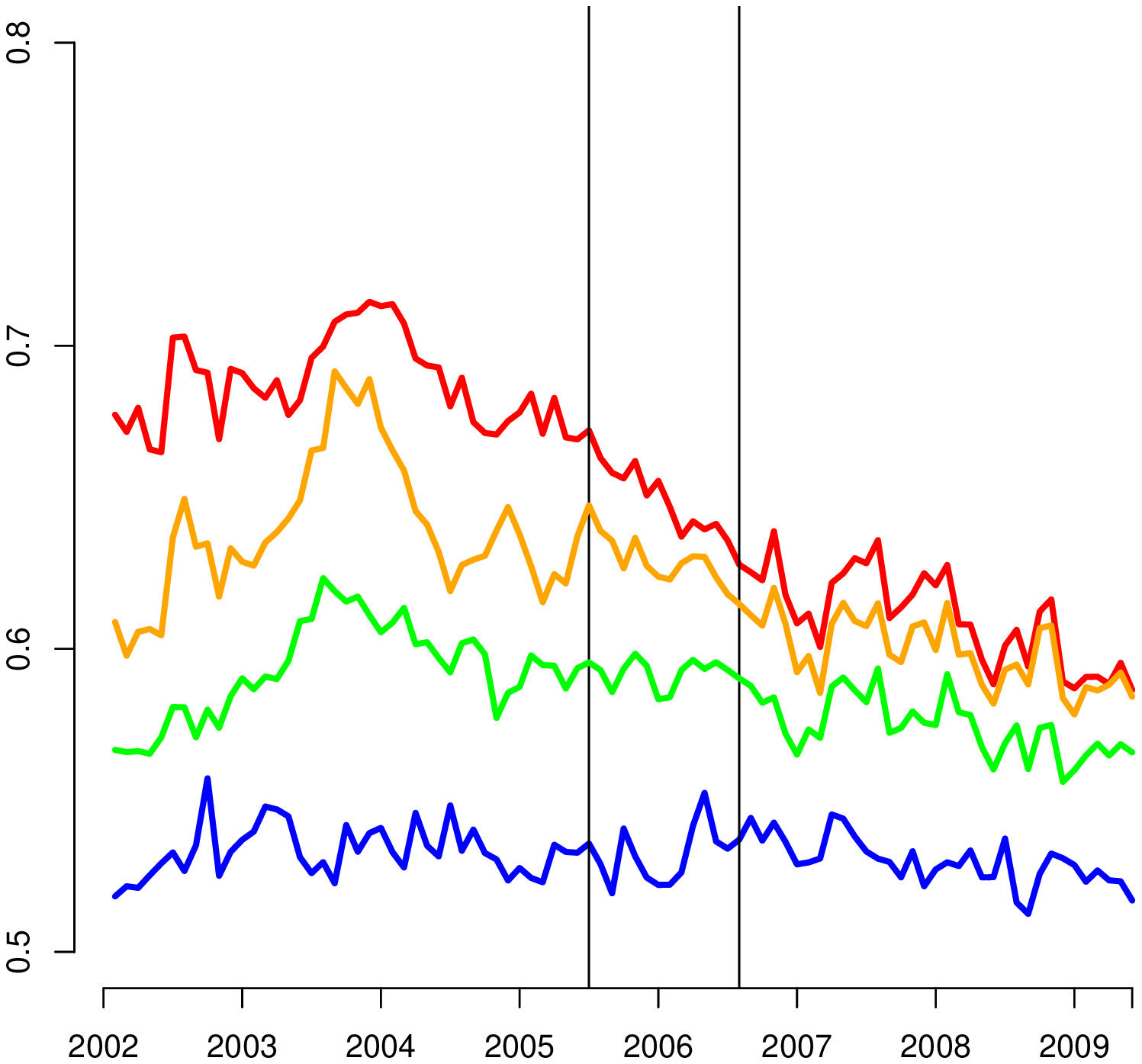}& 
\includegraphics[height=2.75in,width=2.75in]{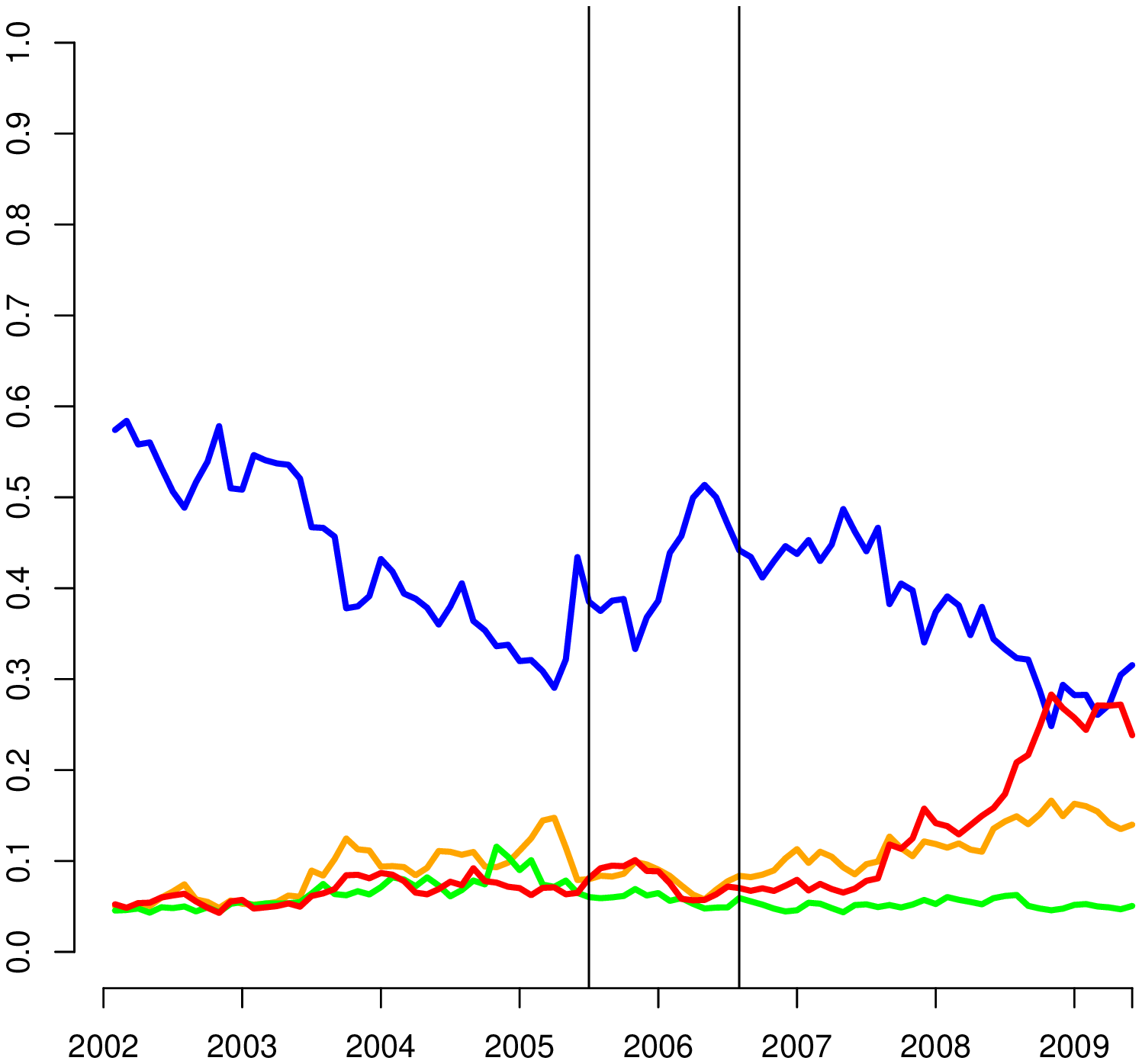}\\
\includegraphics[height=2.75in,width=2.75in]{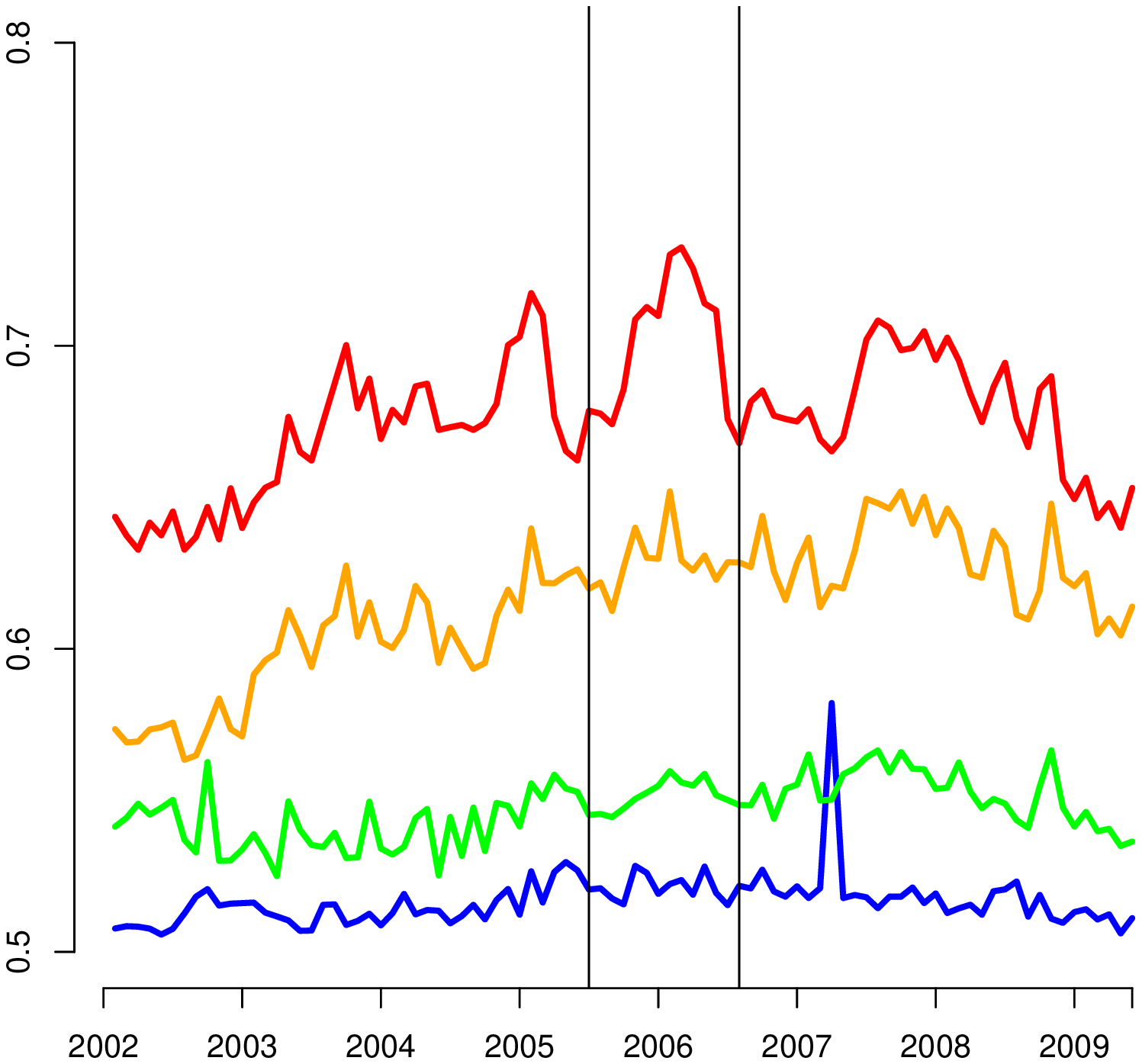}& 
\includegraphics[height=2.75in,width=2.75in]{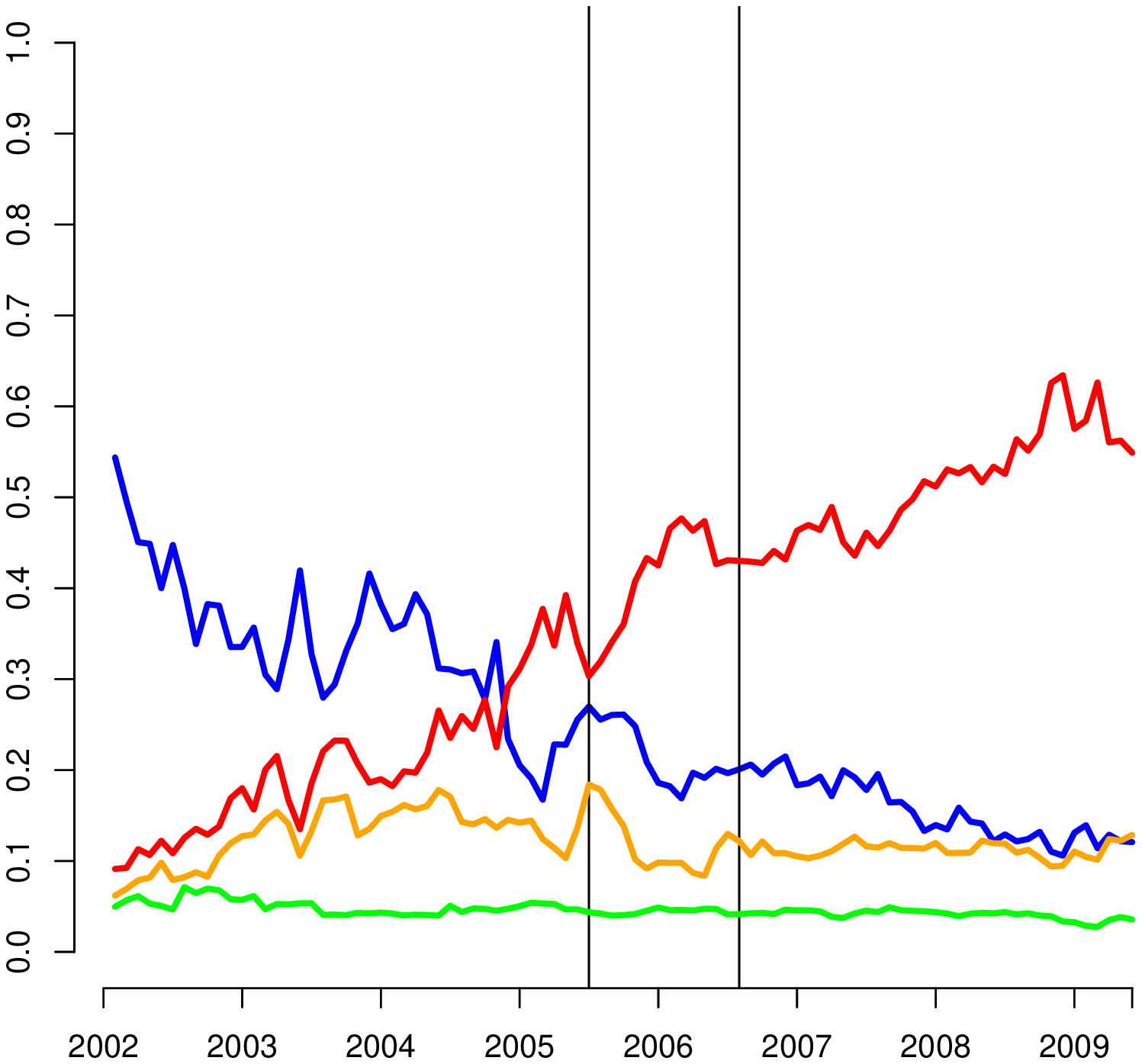}\\
\end{tabular}
\label{3dhurst2}
\caption{Each row of images shows the average Hurst exponent
 by month for a range of trade sizes in the first image. The second image is the relative proportion of all trades by that trade size range. Row by row are Intel Corp. (INTC) and Apple Computer (AAPL). The color lines indicate the range of shares per trade analyzed: blue is 1500+ shares per trade, green is 750-1000 shares per trade, orange is 250-500 shares per trade, and red is less than 250 shares per trade.}
\end{figure}
\pagebreak

In a final check for true self-similarity, the data was analyzed again where the 1s buckets in each day were randomly shuffled. In each case, the correlation structure of the time series was destroyed and $H\approx 0.5$ for all times from 2002-May 2009.

\section{Possible causes of self-similarity}

Given the complex nature of HFT trades and the frequent opacity of firm trading strategies, it is difficult to pinpoint exactly what about HFT causes a higher correlation structure. One answer could be that HFT is the only type of trading that can exhibit trades that are reactive and exhibit feedback effects on short timescales that traditional trading generates over longer timescales.

Another cause may be the nature of HFT strategies themselves. Most HFT strategies can fall into two buckets \cite{CMU}:

\begin{enumerate}
    \item Optimal order execution: trades whose purpose is to break large share size trades into smaller ones for easier execution in the market without affecting market prices and eroding profit. There are two possibilities here. One that the breaking down of large orders to smaller ones approximates a multiplicative cascade which can generate self-similar behavior over time \cite{mandelbrot}. Second, the queuing of chunks of larger orders under an M/G/$\infty$ queue could also generate correlations in the trade flow. However, it is questionable whether the ``service time'', or time to sell shares in a limit order, is a distribution with infinite variance as this queuing model requires.
	\item Statistical arbitrage: trades who use the properties of stock fluctuations and volatility to gain quick profits. Anecdotally, these are most profitable in times of high market volatility. Perhaps since these algorithms work through measuring market fluctuations and reacting on them, a complex system of feedback based trades could generate self-similarity from a variety of yet unknown processes. 
\end{enumerate}
Since firm trade strategies are carefully guarded secrets, it is difficult to tell which of these strategies predominate and induces most of the temporal correlations.

\section{Conclusion}

Given the above research results, we can clearly demonstrate that HFT is having an increasingly large impact on the microstructure of equity trading dynamics. We can determine this through several main pieces of evidence. First, the Hurst exponent $H$ of traded value in short time scales (15 minutes or less) is increasing over time from its previous Gaussian white noise values of 0.5. Second, this increase becomes most marked, especially in the NYSE stocks, following the implementation of Reg NMS by the SEC which led to the boom in HFT. Finally, $H>0.5$ traded value activity is clearly linked with small share trades which are the trades dominated by HFT traffic. In addition, this small share trade activity has grown rapidly as a proportion of all trades. The clear transition to HFT influenced trading noise is more easily seen in the NYSE stocks than with the NASDAQ stocks except NWS. The main exceptions seem to be GENZ and GILD in the NASDAQ which are less widely traded stocks. There are values of $H$ consistently above 0.5 but not to the magnitude of the other stocks. The electronic nature of the NASDAQ market and its earlier adoption of HFT likely has made the higher $H$ values not as recent a development as in the NYSE, but a development nevertheless.

Given the relative burstiness of signals with $H>0.5$ we can also determine that volatility in trading patterns is no longer due to just adverse events but is becoming an increasingly intrinsic part of trading activity. Like internet traffic \cite{bellcore}, if HFT trades are self-similar with $H>0.5$, more participants in the market generate more volatility, not more predictable behavior. The probability of a traded value greater than $V$ in any given time can be given by

\begin{equation}
P[v \geq V] \sim h(v)v^{-\alpha} 
\end{equation}

where $h(v)$ is a function that slowly varies at infinity and $0 < \alpha <2$. The Hurst exponent is related to $\alpha$ by

\begin{equation}
H = \frac{(3-\alpha)}{2}
\end{equation}

There are a few caveats to be recognized. First, given the limited timescale investigated, it is impossible to determine from these results alone what, if any, long-term effects are incorporating the short-term fluctuations.  Second, it is an open questions whether the benefits of liquidity offset the increased volatility. Third, this increased volatility due to self-similarity is not necessarily the cause of several high profile crashes in stock prices such as that of Proctor \& Gamble (PG) on May 6, 2010 or a subsequent jump (which initiated circuit breakers) of the Washington Post (WPO) on June 16, 2010. Dramatic events due to traceable causes such as error or a rogue algorithm are not accounted for in the increased volatility though it does not rule out larger events caused by typical trading activities. Finally, this paper does not investigate any induced correlations, or lack thereof, in pricing and returns on short timescales which is another crucial issue.

Traded value, and by extension trading volume, fluctuations are starting to show self-similarity at increasingly shorter timescales. Values which were once only present on the orders of several hours or days are now commonplace in the timescale of seconds or minutes. It is important that the trading algorithms of HFT traders, as well as those who seek to understand, improve, or regulate HFT realize that the overall structure of trading is influenced in a measurable manner by HFT and that Gaussian noise models of short term trading volume fluctuations likely are increasingly inapplicable.

%Hurst level chart

\appendix
\section{Appendix: Short introduction to wavelets}

Wavelets are a tool to analyze the structure of data series over a variety of scale resolutions. It is often used as an alternative to Fast Fourier Transforms (FFT) in time signals to analyze the frequency responses of signals over various time intervals instead of over the entire signal as in an FFT and is good at revealing sharp spikes or transients FFT would otherwise miss. The reader is encouraged to learn about wavelets in detail using one of the referenced books \cite{wavelet1,wavelet2,wavelet3,wavelet4}, however, a simple overview is given here for assistance in understanding the paper's methodology. 

The basic operation of wavelet analysis is the use of a ``mother wavelet'' $\psi_0$, which has a key feature known as a compact support which means the support is limited to a constrained time period and frequency band. Like Fourier transforms, wavelets can be continuous, or as often used in computer analysis, discrete. There are many mother wavelets, this paper uses the Haar wavelet, but they all share the same basic mathematical properties and $\int^{\infty}_{-\infty} \psi_0(t) dt = 0$. One can use wavelets in a discrete wavelet transform (DWT) to analyze the signal, breaking it down into discrete coefficients, without losing any information about the signal.

The way coefficients are generated is by stretching the wavelet, where each stretch transformation corresponds to an octave, and then by translating the wavelet over each segment of the signal of length $2^{j}\lambda_0$ where $j$ is the octave and $\lambda_0$ is the sampling frequency of the signal. So for a signal with 256 one second readings, the first octave $j=1$ will produce 128 coefficients $(256/2^1)$ and the second octave $j=2$ will have 64 coefficients, etc. There will likely be only 7 octaves analyzed where $j=7$ has only 2 coefficients. The continuous wavelet transform, of which DWT is a discrete version, is shown below and is essentially a convolution of the signal $x(t)$ against a wavelength where the stretch coefficient $a$ and the translation coefficient $b$ determine which part of the signal the wavelet convolves against. $T(a,b)$ is one coefficient corresponding to the stretch of $a$ and the translation of $b$. These coefficients are repeated to account for the entire signal.

\begin{equation}
T(a,b) =
\frac{1}{\sqrt{a}}\int^{\infty}_{-\infty}x(t)\psi_0\left(\frac{t-b}{a}
\right)dt
\end{equation}

To calculate the DWT detailed coefficients $d_{j,k}$ we use

\begin{equation}
d_{j,k} =
\int^{\infty}_{-\infty}x(t)\psi_0\left(\frac{t-k}{2^j}
\right)dt
\end{equation}

The logscale diagram technique used in this paper is based on analyzing the moments of wavelet coefficients. The $n$th moment of a collection of wavelet coefficients in an octave is given by
\begin{equation}
S_n(j) = \frac{1}{n_j}\sum_k |d_X(j,k)|^n
\end{equation}

Most commonly, the second moment is used but other moments can be used in a similar manner to gain more information about the signal. The slope of $\log S_2(j)$ vs. $j$ is related to the Hurst exponent over a range of octaves $j\in [j_1,j_2]$ by the equation

\begin{equation}
S_2(j)=c|2^{-j}\lambda_0|^{1-2H}
\end{equation}

where $c$ is a constant and $\lambda_0$ is the sampling frequency of the signal. A logarithm of base 2 must be used for correct calculations. So

\begin{equation}
\log S_2(j)=c_2(2H-1)j
\end{equation}

where $c_2$ is a new constant combining $c$ and the sampling frequency.

The use of this method across many octaves was widely used in the investigation of Internet traffic dynamics in the late 1990s and early 2000s. A rigorous overview of the method is given in \cite{hurstguide1,hurstguide2}


\begin{harvard}
\bibitem[Abry et. al. (2000)]{hurstguide1}Abry, P., Flandrin, P., Taqqu, M.S. \& Veitch, D. Wavelets for the analysis, estimation, and synthesis of scaling data. In \emph{Self-similar network traffic and performance evaluation} eds. Park, K. \& Willinger, W.  2000 (John Wiley \& Sons: New York) 39-88. 
\bibitem[Abry et. al. (2003)]{hurstguide2}Abry, P., Flandrin, P., Taqqu, M.S. \& Veitch, D. Self-similarity and long-range dependence through the wavelet lens. In \emph{Theory and Applications of Long-Range Dependence} eds. Doukan, P., Taqqu, M.S., Oppenheim, G. 2003, (Birkh\"{a}user: Boston) 527-556.
\bibitem[Addison (2002)]{wavelet4}Addison, P. The Illustrated Wavelet Transform
Handbook, 2002 (CRC Press: Boca Raton).
\bibitem[Bardet and Kammoun (2008)]{DFAcriticize}Bardet, J.M. \& Kammoun, I. Asymptotic Properties of the Detrended Fluctuation Analysis of Long Range Dependent Processes \emph{IEEE Transactions on Information Theory}, 2008, \textbf{54}, 2041-2052.
\bibitem[Couillard and Davison (2005)]{RScriticize}Couillard, M. \& Davison, M. A comment on measuring the Hurst exponent of financial time series. \emph{Physica A}, 2005, \textbf{348} 404–418.
\bibitem[Degryse et. al. (2008)]{mktinfo5}Degryse, H., van Achter, M., and Wuyts, G. Shedding Light on Dark Liquidity Pools.\emph{TILEC Discussion Paper} DP 2008-039, 2008.
\bibitem[Eisler et. al. (2005)]{intraday1}Eisler, Z., Kert\'{e}sz, J., Yook, S.H., \& Barab\'{a}si, A.L., Multiscaling and non-universality in fluctuations of driven complex systems. \emph{Europhysics Letters}, 2005, \textbf{69}, 664-670.
\bibitem[Eisler and Kert\'{e}sz (2007a)]{intraday2}Eisler, Z. \& Kert\'{e}sz, J., Liquidity and the multiscaling properties of the volume traded on the stock market. \emph{Europhysics Letters}, 2007, \textbf{77}, 28001.
\bibitem[Eisler and Kert\'{e}sz (2007b)]{intraday3}Eisler, Z. \& Kert\'{e}sz, J., The dynamics of traded value revisited. \emph{Physica A}, 2007, \textbf{382}, 66-72.
\bibitem[Francis et. al. (2009)]{mktinfo3}Francis, J. C., Harel, A. and Harpaz, G.  Exchange Mergers and Electronic Trading. \emph{The Journal of Trading}, 2009, \textbf{4}:1 35-43.
\bibitem[Grau-Carles (2005)]{price3}Grau-Carles, P. Tests of Long Memory: A Bootstrap Approach. \emph{Computational Economics}, 2005, \textbf{25} 103-113.
\bibitem[Harris (2003)]{markethistory2}Harris, L. Trading and Exchanges: Market microstructure for practitioners. 2003, (Oxford: New York).
\bibitem[Kaiser (1994)]{wavelet3}Kaiser, G. A Friendly Guide to Wavelets, 1994, (Springer: Berlin).
\bibitem[Lehoczky and Schervish (2009)]{CMU}Lehoczky, J. and Schervish, M.  Ch. 9 High Frequency Trading (lecture notes from Carnegie Mellon University, Pittsburgh, PA) 2009.
\bibitem[Leland et. al. (1994)]{bellcore}Leland, W.E., Taqqu, M.S., Willinger, W. and Wilson, D.V. On the self-similar nature of Ethernet traffic (extended version). \emph{IEEE/ACM Transactions on Networking}, 1994, \textbf{2}:1, 125-151.
\bibitem[Liesenfeld (2002)]{volume2}Liesenfeld, R. Identifying common long-range dependence in volume and volatility using high frequency data. SSRN paper ID: 326300, 2002.
\bibitem[Lo (1991)]{price1}Lo, A.W., Long-term memory in stock market prices. \emph{Econometrica}, 1991, \textbf{59}:5 1279-1313.
\bibitem[Lobato and Savin (1998)]{price2}Lobato, I.N. and Savin, N.E. Real and Spurious Long-Memory Properties of Stock-Market Data. \emph{Journal of Business \& Economic Statistics}, 1998, \textbf{16}:3 261-268.
\bibitem[Lobato and Velasco (2000)]{volume1}Lobato I.N. and Velasco, C. Long memory in stock-market trading volume. \emph{Journal of Business \& Economic Statistics}, 2000, \textbf{18}:4 410-427.
\bibitem[Mandelbrot (1974)]{mandelbrot}Mandelbrot, B.B. Intermittent turbulence in self-similar cascades: Divergence of high moments and dimension of the carrier." \emph{Journal of Fluid Mechanics}, 1974, \textbf{62} 331-358.
\bibitem[Mantegna and Stanley (1999)]{econo1}Mantegna, R.N. and Stanley, H.E. An
Introduction to Econophysics: Correlations and Complexity in
Finance, 1999 (Cambridge University Press: Cambridge)
\bibitem[Markham and Harty (2008)]{markethistory1} Markham, J.W. and Harty, D.J.  For Whom the Bell Tolls: The Demise of Exchange Trading Floors and the Growth of ECNs. \emph{Journal of Corporation Law}, 2008, \textbf{33}:4 866-939.
\bibitem[McAndrews and Stefandis (2000)]{mktinfo2}McAndrews, J. and Stefanadis, C. The Emergence of Electronic Communications Networks in the U.S. Equity Markets. \emph{Current Issues in Economics and Finance (Federal Reserve Bank of New York)}, 2000, \textbf{6}:12 1-6
\bibitem[Mittal (2008)]{mktinfo4}Mittal, H. Are You Playing in a Toxic Dark Pool? A Guide to Preventing Information Leakage. \emph{The Journal of Trading}, 2008, \textbf{3}:3 20-33.
\bibitem[Nievergelt (1999)]{wavelet2}Nievergelt, Y. Wavelets Made Easy, 1999, (Springer: Berlin)
\bibitem[Palmer (2009)]{mktinfo6}Palmer, M. Algorithmic Trading: A Primer. \emph{The Journal of Trading}, 2009, \textbf{4}:3 30-35.
\bibitem[Percival and Walden (2000)]{wavelet1}Percival, D.B. and Walden, A.T. Wavelet Methods for Time Series Analysis, 2000, (Cambridge University Press: New York)
\bibitem[Stoll (2006)]{mktinfo1}Stoll, H.R. Trading in Stock Markets. \emph{The Journal of Economic Perspectives}, 2006, \textbf{20}:1 153-174
\bibitem[Tsay (2002)]{econo2}Tsay, R.S. Analysis of Financial Time Series. 2002. (Wiley: New York).
\end{harvard}
\end{document}